\documentclass[letterpaper,twocolumn,10pt]{article}
\usepackage{usenix,epsfig} 
\usepackage{subfig}
\usepackage{graphicx}

\usepackage{amsmath}
\usepackage{amssymb}

\newcommand{\ts}{Okapi}

\newcommand{\remove}[1]{}

\usepackage{algpseudocode}
\usepackage{algorithm}
\usepackage{cancel}
\usepackage{bm} 
\usepackage{xcolor} 
\usepackage{enumerate}
\usepackage{enumitem} 
\usepackage{amsthm} 
\usepackage{verbatim} 

\algdef{SE}[DOWHILE]{Do}{doWhile}{\algorithmicdo}[1]{\algorithmicwhile\ #1}%

\newcommand{\pvs}{\vspace{-9pt}}
\newcommand{\param}[1]{{\bf$\langle$#1$\rangle$}} 
\newcommand{\la}{$\leftarrow$}
\newcommand{\lam}{\gets}
\newcommand{\mib}[1]{$\bm{#1}$}
\newcommand{\algsize}{\scriptsize} 
\renewcommand{\Comment}[1] {\hfill\textit{\textcolor{cerulean}{$\triangleright$~#1}}} 

\algdef{SE}[EVENT]{Event}{EndEvent}[1]{\textbf{upon}\ #1\ \algorithmicdo}{\algorithmicend\ \textbf{event}}%
\algtext*{EndEvent}

\makeatletter
\algnewcommand{\OnlyComment}[1] {\hskip\ALG@thistlm\textit{\textcolor{cerulean}{$\triangleright$~#1}}}
\algnewcommand{\LineIndentComment}[1] {\Statex \hskip\ALG@thistlm\hskip\algorithmicindent\textit{\textcolor{cerulean}{$\triangleright$~#1}}} 
\algnewcommand{\LineComment}[1] {\Statex \hskip\ALG@thistlm\textit{\textcolor{cerulean}{$\triangleright$~#1}}} 
\makeatother

\algnewcommand\algorithmicforeach{\textbf{for each}}
\algdef{S}[FOR]{ForEach}[1]{\algorithmicforeach\ #1\ \algorithmicdo}

\algnewcommand{\EndIIf}{\unskip\ \algorithmicend\ \algorithmicif}
\algrenewcommand\algorithmicindent{1em}
\definecolor{cerulean}{rgb}{0.0, 0.48, 0.65}
\usepackage{array}
\usepackage{calc}

\newtheorem{observation}{Observation}
\newtheorem{proposition}{Proposition}
\begin{document}

\date{}

\title{\Large \bf \ts{} : Causally Consistent Geo-Replication Made Faster, Cheaper\\ {\color{black}and More Available}}

\author{
{\rm Diego Didona, Kristina Spirovska, Willy Zwaenepoel}\\
\'Ecole polytechnique f\'ed\'erale de Lausanne
} 

\maketitle

\thispagestyle{empty}

\subsection*{Abstract}
\ts{} is a new causally consistent geo-replicated key-value store. \ts{} leverages two key design choices to achieve high performance. First, it relies on hybrid logical/physical clocks to achieve low latency even in the presence of clock skew. 
 Second, \ts{} achieves higher resource efficiency and better availability, at the expense of a slight increase in update visibility latency. To this end, \ts{} implements a new stabilization protocol that uses a combination of vector and scalar clocks and makes a remote update visible when its delivery has been acknowledged by every data center. 


We evaluate \ts{} with different workloads on Amazon AWS, using three geographically distributed regions and 96 nodes. We compare \ts{} with two recent approaches to causal consistency, Cure and GentleRain. 
We show that \ts{} delivers up to two orders of magnitude better performance than GentleRain 
 and that \ts{} achieves up to 3.5x lower latency and a 60\% reduction of the meta-data overhead with respect to Cure.

\section{Introduction}
\label{sec:introduction}
Distributed data stores represent the backbone of many large-scale online services. Such data stores are often geo-replicated to improve performance, by storing a copy of the data closer to the clients, and to achieve availability, by keeping multiple copies of the data at different sites~\cite{Nishtala:2013}. A critical decision in designing a geo-replicated store is the choice of its consistency model. 
At one end of the spectrum, strong consistency~\cite{Herlihy:1990} has simple semantics, but incurs high latency and does not tolerate network partitions. At the other end, eventual consistency provides excellent performance and tolerates partitions~\cite{Vogels:2009}, but it is hard to program with.

\pvs
~~\\\noindent{\bf Causal Consistency.} Causal consistency~\cite{Ahamad:1995} hits a sweet spot in the ease of programming vs performance trade-off and has emerged as an attractive model to build geo-replicated data stores~\cite{Akkoorath:2016,Almeida:2013,Bailis:2013,Du:2014,Lloyd:2011}.
\remove{ and is an attractive model for building geo-replicated data stores. 
Causal consistency has garnered much attention as it hits a sweet spot in the ease of programming vs performance trade-off for this kind of systems~\cite{Akkoorath:2016,Almeida:2013,Bailis:2013,Du:2014,Lloyd:2011}.} 
 On the one hand, it avoids the long latencies and inability to tolerate network partitions of strong consistency. On the other hand, it is easier to reason about than eventual consistency and avoids some of eventual consistency anomalies.

\pvs
~~\\\noindent{\bf Limitations of existing systems.} 
At a very high level, all causally consistent systems work in the same way. Events, such as creation, reads and writes of data items, are labeled with a timestamp. This timestamp is propagated on communications between machines. New events are then always labeled with timestamps higher than the highest one received so far, thereby making sure that timestamps reflect causal order.
A variety of timestamping methods have been proposed, including in particular using the current value of the physical clock of the machine on which the event occurs~\cite{Akkoorath:2016,Du:2014}. The advantage of using physical clocks is that it is a rather concise encoding of causality, and that it is trivial to obtain. 

The problem with using physical clocks for causal dependency tracking is due to clock skew between different machines. Consider, for instance, the creation of a new version of data item X on machine A, occurring at time $t$ on the physical clock of A, and therefore labeled with timestamp $t$. Let this version of X be read by a client on machine B, and let that client then create  a new version of data item Y. 
In the absence of clock skew, the value $t'$ of B's clock at this time would be larger than $t$. Hence, this event can 
 be timestamped with $t'$, with the timestamp reflecting the causal order.
However, if, due to clock skew, the value of B's clock at this time is smaller than $t$, then the operation needs to block until B's clock catches up to the value of $t$~\cite{Du:2014}. 

Clock skew yields similar problems also with read-only transactions, a powerful abstraction supported by the majority of the causally consistent systems~\cite{Akkoorath:2016,Almeida:2013,Du:2014,Lloyd:2011,Lloyd:2013}. The timestamp assigned to a transaction, in fact, can be higher than the value of the clock on a node involved in the transaction~\cite{Akkoorath:2016}.
Recent work has shown that waits due to clock skew cause a significant reduction in performance (up to 25\% even in a small deployment~\cite{Akkoorath:2016}).


State-of-the-art systems based on physical clocks also make different trade-offs between dependency tracking overhead and the latency of transactional operations. 
They either use a single timestamp per item but achieve poor performance for transactional operations~\cite{Du:2013}, or use a number of timestamps equal to the number of data centers to implement efficient transactions~\cite{Akkoorath:2016}.

Modern applications, however, need {\em both} fast transactional read operations on a consistent snapshot of the data store {\em and} low dependency tracking overhead. Multi-key transactional reads are paramount, as they increase the expressiveness of applications~\cite{Lloyd:2011,Lu:2016}.  For example, in many services a single high-level user operation (e.g., retrieving the content of a page) translates to multiple read operations from the underlying store~\cite{Nishtala:2013}. It is, then, highly desirable that all the retrieved values belong to a consistent causal snapshot of the data store. Moreover, small items dominate typical workloads, e.g., at Facebook~\cite{Atikoglu:2012,Nishtala:2013}, Twitter~\cite{twitter} and Instagram~\cite{instagram}. For such workloads, dependency meta-data can easily grow bigger than the payload it refers to, with detrimental effects on scalability, communication and storage overhead. 
 
 \pvs
 ~\\\noindent{\bf \ts{}.} This paper presents \ts{}, a new causally consistent geo-replicated data store. \ts{} avoids the latencies caused by clock skew and implements efficient transactional reads at low dependency tracking cost.
 
\ts{} achieves the first goal by using hybrid logical/physical clocks (HLC) to timestamp events. HLC have a physical and a logical component.
Instead of waiting for the physical clock to reach a value $t$,  a server can set the physical component of its hybrid clock to $t$. The logical part of the clock is, then, incremented to generate new timestamps that reflect causality among events.
 
To achieve the second goal, \ts{} proposes a novel stabilization protocol called Universal Stable Time (UST). UST achieves higher resource efficiency and better availability, at the expense of a slight increase in update visibility latency. UST uses dependency vectors only for local updates, to efficiently serve transactional reads, and a single timestamp for replicated updates.  UST provides support for higher availability during network partitions, by enforcing that data centers expose to clients only items that have been received system-wide. The (periodic and asynchronous) communication needed to check the set of remotely received items induces a slightly higher visibility latency for remote updates.

\pvs
~\\\noindent{\bf Contributions.} This paper makes three contributions:

\pvs
~\\\noindent{\bf I) } The design and implementation of \ts{}, a causally consistent data store that $i)$ achieves low latencies by means of a novel combination of HLC and dependency vectors; and $ii)$ proposes a novel stabilization protocol that achieves higher efficiency and availability at the cost of a slight increase in remote updates visibility latency.

\pvs
~\\\noindent{\bf II) }The exploration of \ts{}'s trade-offs among performance, updates visibility latency and availability. 

\pvs
~\\\noindent{\bf III) } The evaluation of \ts{} in a large scale Amazon AWS deployment, in which we compare \ts{} with Cure and GentleRain, two state-of-the-art systems that achieve causal consistency using physical clocks.

\remove{The evaluation of \ts{} in a large scale Amazon AWS deployment, using up to 96 nodes scattered across up to 3 data centers, against two state-of-the-art approaches to causal consistency based on scalar and vector physical clocks. Such evaluation shows that \ts{} delivers up to one order of magnitude better performance than the approach based on scalar clock and that \ts{}  achieves up to {\color{red} xx\%} lower latency and a {\color{red}yy\%} reduction of the meta-data overhead withe respect to the approach based on vector clocks. {\color{blue}Avoid details?}
}

\remove{
 $i)$ \ts{} delivers up to one order of magnitude better performance with transactional workloads; $ii)$ \ts{} reduces dependencies tracking meta-data by up to a factor of 2 when compared to the state-of-the-art approach to transactional causal consistency; and $iii)$ \ts{} is competitive with existing systems with workloads composed by single read and write operations.
{\color{blue}remove the competitive stuff. We are only going to evaluate a mixed workload now, where we hide the fact that get operations are more efficient in GentleRain}
}
\remove{
{\color{red}The remainder of the paper is organized as follows. Section~\ref{sec:def} describes the target system model. Section~\ref{sec:design} presents the design of \ts{}. Section~\ref{sec:protocol} describes the protocols in \ts{}. Section~\ref{sec:corr} provides correctness arguments. Section~\ref{sec:comp} qualitatively compares \ts{} with existing solutions. Section~\ref{sec:eval} describes the quantitative evaluation and comparison of \ts{} against state-of-the-art systems. Section~\ref{sec:rw} discusses related work. Finally, Section~\ref{sec:concl} concludes the paper.}
}
\section{Definitions and System model}
\label{sec:def}
{\bf Causal consistency.} Causal consistency requires that servers of a system return values that are consistent with the order defined by the {\em causality} relationship.
Causality is a happens-before relationship between two events~\cite{Lamport:1978,Ahamad:1995}. For two operations $a,\  b$, we say that $a$ causally depends on $b$, and write $a \leadsto b$, if and only if at least one of the following conditions holds: $i)$ $a$ and $b$ are operations in a single thread of execution, and $a$ happens before $b$; $ii)$ $a$ is a write operation, $b$ is a read operation, and $b$ reads the value written by $a$; $iii)$ there is some other operation $c$ such that $a \leadsto c$ and $c \leadsto b$.

\remove{Unless stated otherwise, }We use lower case letters, e.g., $x$, to refer to a key and the corresponding capital letter, e.g., $X$ to refer to a version of the key. We say that $X$ depends on $Y$ if the write of $X$ causally depends on the write of $Y$.


We define an item {\em stable} in a data center when it becomes visible to clients in that data center. An item becomes stable when all its dependencies have been received and made visible in the local data center. 

We define the {\em visibility latency} of an item $d$ in a data center $DC$ as the time between the moment in which $d$ has been created in its originating data center and the moment in which $d$ becomes stable in $DC$.

\pvs
~\\\noindent{\bf Convergent conflict handling.} Two operations $a,\ b$ are {\em concurrent} if neither $a\leadsto b$ nor $b\leadsto a$. If $a$ and $b$ are concurrent write operations to the same key, they {\em conflict}. Two conflicting versions of a key can be propagated to remote replicas in different orders, potentially leading to replicas to diverge forever. \ts{} implements the popular last-writer-wins rule~\cite{Thomas:1979} to arbitrate conflicting modifications to keys. Given two updates, the one with the highest timestamp is deterministically decided to having occurred later than the other, determining the value of the value written (possible ties are settled by looking at the id of the originating data centers of the items). \ts{} can easily integrate other mechanisms to achieve state convergence, similarly to previous systems~\cite{Akkoorath:2016,Du:2013,Du:2014,Lloyd:2011}.  
\remove{
The causality relationship only determines a partial order among operations. Two operations $a,\ b$ are concurrent if neither $a\leadsto b$ nor $b\leadsto a$. If both $a$ and $b$ are concurrent write operations towards the same key, they {\em conflict}. Two conflicting versions of the same key can be propagated to remote replicas in different orders, potentially leading to replicas to diverge forever. 
To enforce that different replicas converge to the same value for any given key, a system must implement a convergent conflict handling procedure. Such procedure must implement a commutative and associative function, so that replicas can manage update replication messages in the order they receive them and converge to the same state.

The most popular convergent conflict handling functions relies on the last-writer-wins rule~\cite{Thomas:1979}. Given two updates, one of them is deterministically decided to having occurred later than the other, determining the value of the value written. 
For simplicity, \ts{} implements the last-writer-wins rule based on the timestamp of items (possible ties are settled by looking at the id of the originating data centers of items), but it could support other mechanisms similarly to previous systems~\cite{Lloyd:2011,Du:2013,Du:2014,Akkoorath:2016}. 
}
 
 \pvs
~\\\noindent{\bf System model.}  
We assume a distributed key-value store that manages a large set of data items. The data-set is split into $N$ partitions and each key is deterministically assigned to one partition according to a hash function. Each partition is replicated at $M$ different sites, each corresponding to a different data center. Hence, a full copy of the data is stored at each site.

We assume a multiversion data store. An update operation creates a new version of a key. Each version stores the value corresponding to the key and some meta-data  to track causality. The system periodically garbage-collects old versions of keys. 
We further assume nodes communicate through point-to-point lossless FIFO channels.

The system supports the same programming model of the vast majority of the existing causally consistent systems, e.g., COPS~\cite{Lloyd:2011}, Orbe~\cite{Du:2013}, ChainReaction~\cite{Almeida:2013}, and Gentlerain~\cite{Du:2014}, which is based on these operations:

\pvs
~\\\noindent{\bf PUT(key, val): } A PUT operation assigns value $val$ to an item identified by $key$. If item $key$ does not exist, the system creates a new item with initial value $val$. Else, a new version storing $val$ is created.

\pvs
~\\\noindent{\bf val $\gets$ GET(key): } A GET operation returns the value of the item identified by $key$. A GET operation is such that its return value does not break causal consistency as explained in the following. Assume $X \leadsto Y$ and that a client $c$ issues a GET($y$) operation, receiving $Y$ as result. Then, any subsequent GET($x$) operation issued by $c$ must return either $X$ or a version $X'$ such that $X' \cancel{\leadsto} X$.

\pvs
~\\\noindent{\bf $\langle$vals$\rangle\ \gets$ RO-TX $\langle$keys$\rangle$: } This operation implements a causally consistent read-only transaction~\cite{Lloyd:2011,Lloyd:2013}.  If a read-only transaction returns $X$ and $Y$, then they are causally consistent with the issuing client's history and there is no $X'$, such that $X\leadsto X'\leadsto Y$.

At the beginning of a session, a client $c$ connects to a node $n_c$ in the closest data center according to some load balancing scheme. $c$ does not issue the next operation until it receives the reply to the current one. Operations towards data items that are not stored by $n_c$ are transparently forwarded to the node(s) responsible for such data items, and the result is relayed back to $c$ by $n_c$.

\begin{figure}[t]
\centering
\includegraphics[scale=0.2]{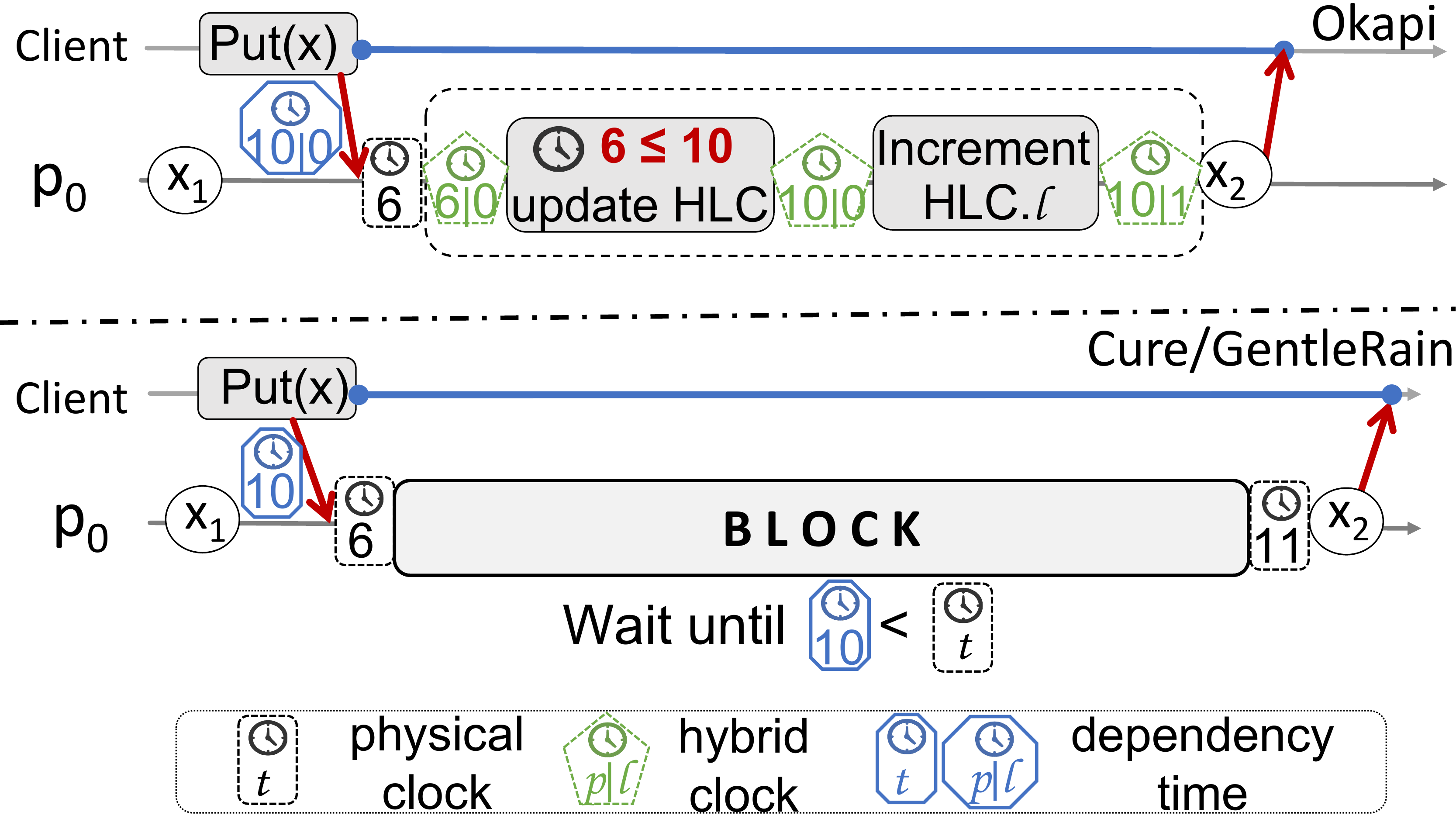}
\caption{PUT implementation in \ts{} (top) and GentleRain/Cure (bottom). The client dependency time (10) is higher than the physical clock on $p_0$ (6). To reflect causality, \ts{} sets its HLC to $\langle 10,1\rangle$. GentleRain/Cure must wait until the physical clock gets to 11.}
\label{fig:hyb}
\end{figure}

Each server is equipped with a physical clock that advances monotonically. We assume such clocks to be loosely synchronized by a time synchronization protocol, such as NTP~\cite{ntp}. The correctness of our protocol does not depend on the synchronization precision.
\section{The Design of \ts{}}
\label{sec:design}
We now describe the design of \ts{}, focusing on its two key techniques: the use of HLC  and the UST stabilization protocol. We also qualitatively compare \ts{} with two state-of-the-art systems, GentleRain and Cure~\footnote{Cure exposes APIs different from \ts{}'s and uses CRDTs~\cite{Shapiro:2011} for state convergence. We have implemented a version of Cure that complies with the system model described in Section~\ref{sec:def}. We refer to our implementation simply as Cure.}.

\remove{
\subsection{Overview}
\label{sec:design:ovw}
Each item $d$ in \ts{} is timestamped with an {\em update time}, which represents the time measured on the server $p$ when $p$ created $d$. $d$ also carries an {\em item dependency vector}, with one entry per data center. This vector summarizes the dependencies of $d$. The local entry of the vector corresponds to the update time of $d$. The entries corresponding to remote data centers, instead, track the remote dependencies of $d$. Let $c$ be the client that has created $d$. Then, the (remote) entry with index $i$ of $d$'s dependency the vector represents the highest timestamp among the ones of the items created at data center $i$ on which $c$ depended when creating $d$.  

Clients keep track of established dependencies by means of a {\em client dependency vector}, with one entry per data center. The semantics of this vector is the same as the item dependency vector's. The client dependency vector is provided upon performing any operation. In the RO-TX and GET case, this is done to ensure that the operation is served from a snapshot that includes the dependencies established by the client.
 In the PUT case, the vector is provided with a twofold goal. First, it is used to build the newly created item's dependency vector. Second, it allows the receiving server to enforce that the update time of the new item is higher than the highest timestamp seen by the client. In this way, \ts{} maintains the invariant that if $X \leadsto Y$, then the update timestamp of $X$ is lower than $Y$'s.
}
\subsection{Using HLC to track time}
\label{sec:design:hyb}
\ts{} uses HLC~\cite{Kulkarni:2014} to track the advancement of time, and hence to timestamp updates.
A hybrid timestamp $t$ is a tuple with a physical component $t.p$ and a logical component $t.l$. Two hybrid timestamps are compared by first comparing their physical components, and then comparing their logical components.

Each server $p$ has a (software maintained) hybrid machine clock $HLC_p$ and a (hardware maintained) physical clock $Clock_p$. The physical component of the $HLC_p$ is in general different from the current value of $Clock_p$.

Each data item version stored on $p$ has a (hybrid) update timestamp. At the time of creation of a version, its update timestamp is set to the current value of $HLC_p$.

Each client $c$ has a client dependency vector $DV_c$, with one entry per data center. 
  $DV_c$ consists of hybrid timestamps that, roughly speaking, reflect the client's dependencies on data items created at each other data center.

We now show how \ts{} leverages HLC to implement clock-skew resilient PUT and RO-TX operations.

\pvs
~\\\noindent{\bf PUT.} When client $c$ performs a $PUT$ on server $p$, it sends its $DV_c$ along. The server $p$ then computes the largest element of $DV_c$, noted $max_c$, and ensures that the update timestamp of the newly created version is higher than $max_c$ and higher than the highest update timestamp $p$ has assigned so far. In this way, the generated timestamp reflects causality. 

To this end, the server first sets $HLC_p.p$ to the maximum of $HLC_p.p$ and $Clock_p$.  If $max_c < HLC_p$, then $HLC_p.l$ is incremented by one, so that the new update timestamp is higher than any previous one assigned by $p$. Otherwise, $HLC$ is set to $<max_c.p, max_c.l + 1>$, to ensure that the new update timestamp is higher than the highest dependency timestamp of the client. This ensures, without waiting, that the new update timestamp reflects causality.  The server also attaches an item dependency vector to the new item. Such dependency vector is a copy of $DV_c$ except for the entry corresponding to the local data center, which stores the timestamp of the item.

In contrast, if timestamping is done using a physical clock alone, as in Cure and Gentlerain, then in the case of clock skew, the server has no other option but to wait until the physical clock catches up with $max_c$. 

Figure~\ref{fig:hyb} (top) and Figure~\ref{fig:hyb} (bottom) depict the behavior of \ts{} and, respectively, Cure and GentleRain when a node $p$ with $Clock_p = 6$ receives a PUT operation from a client with $max_c = 10$.

\begin{figure}[t!]
\centering
\includegraphics[scale=0.35]{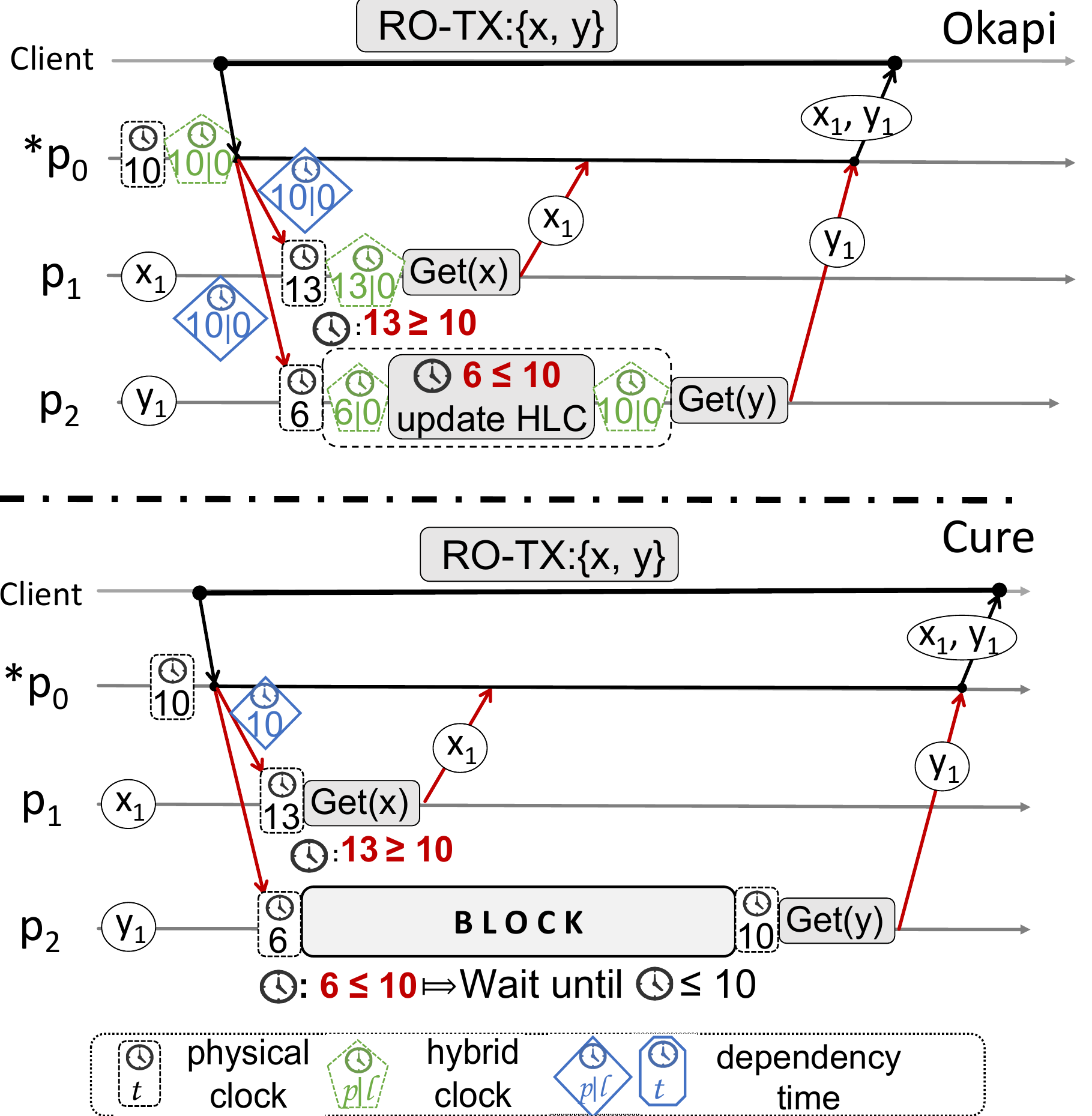}
\caption{RO-TX implementation in \ts{} (top) and Cure (bottom). The local snapshot time of the transaction (10) is higher than the value of the physical clock on $p_2$ (6). To avoid the creation of later items with a timestamp still $\leq 10$, \ts{} simply moves its $HLC$ to $\langle 10,0\rangle$. Cure needs to wait until the clock gets to 10. }
\label{fig:phys}
\end{figure}

\pvs
~\\\noindent{\bf RO-TX.} The advantages of HLC are even greater for read-only transactions.
In this case the transaction coordinator computes a transaction snapshot time, essentially the upper bound on timestamps corresponding to local items that are visible to the transaction~\footnote{The coordinator also determines upper bounds for the remote dependencies, as we shall discuss in the next section. We omit them here for simplicity, as they are not affected by clock skew and do not induce any waiting time in \ts{} and Cure.}.  

The coordinator then sends requests to all the servers storing data items requested in the transaction, asking them to return the values of the versions of those data items with the largest timestamp smaller than or equal to the snapshot time.

Intuitively, when using hybrid timestamps, the server can respond immediately, regardless of clock skew, by if necessary advancing its $HLC$ to the transaction snapshot time. This disallows ``later" items from being created by the server with a timestamp smaller than or equal to the snapshot time, thereby preventing the transaction from ``missing" any item that it should be able to access.

If, instead, physical clocks are used, then the server has no other option than to wait for the clock to catch up to the snapshot time.

The benefits of hybrid clocks are more pronounced with transactions because with physical clocks it suffices that the clock of any of the contacted servers runs behind for the waiting to occur. This easily results in a major performance impairment for systems based only on physical clocks because high-level application operations typically  translate to  contacting several servers at once. For example, the median number of servers contacted to retrieve a Facebook page is  about $20$~\cite{Nishtala:2013}.

Figure~\ref{fig:phys} (top) and Figure~\ref{fig:phys} (bottom) compare the behavior of \ts{} and Cure when serving a RO-TX with snapshot time 10 and with a contacted server $p_2$ whose physical clock value is 6. We only compare \ts{} with Cure because the implementation of RO-TX in Cure is more efficient than in GentleRain. We shall discuss the limitations of the RO-TX implementation in GentleRain in the following section.
\subsection{Efficient dependency tracking by UST}
\label{sec:design:dep}
\ts{} incorporates UST, a new stabilization protocol that addresses availability issues in state-of-the-art causally consistent systems. As a by-product, it considerably reduces the amount of consistency meta-data that is communicated and stored, compared to Cure, approximating that required in GentleRain. As a tradeoff, Okapi incurs a modestly higher update visibility latency.

\pvs
~\\\noindent{\bf UST in a nutshell.} As with most causally consistent systems, UST allows updates originating in a data center to become visible immediately in that data center. For updates originating elsewhere, it makes them visible only when they have been replicated at all data centers. 

UST works by a combination of version vectors on each server that record the latest remote updates received from their replicas in other data centers and a decentralized protocol for exchanging this information to determine what data items are fully replicated. 
 Periodically, nodes within a data center exchange their version vectors to compute the Global Stable Vector (GSV) as the entry-wise minimum of all the version vectors in the data center. If the $i$-th entry of the GSV takes the value $t$, it means that  all the servers in the data center have installed all updates originated at data center $i$ with timestamp up to $t$.  Periodically, peer replicas exchange their GSV to compute the Universal Stable Vector (USV), as entry-wise minimum of all the exchanged GSV. If the $i$-th entry of the USV takes the value $t$, then all updates originated at data center $i$ have been fully replicated.

\pvs
~\\\noindent{\bf Availability.}  In Cure and GentleRain the failure or disconnection of a data center can cause the states of healthy data centers to diverge. Namely, it can happen that some data items originating at that failed data center are visible in some healthy data centers but not in other ones. 
UST disallows this behavior by making a remote data item visible only when it has been replicated at every data center. This ensures that healthy data centers have made visible the same set of items from the failed data center {\color{black} and hence see the same set of remote stable dependencies even after the failure}.



\pvs
~\\\noindent{\bf Meta-data overhead.} UST only requires a single scalar value to be communicated and stored with a remote update {\color{black} to determine its visibility. In fact, when a remote update $d$ coming from data center $i$ arrives in data center $j$, all the remote dependencies of $d$ have been already fully replicated. Hence, UST determines the visibility of $d$ by only checking that all of $d$'s local (i.e., of data center $i$) dependencies have been fully replicated. This is accomplished by simply checking if the $i$-th entry of the USV is lower or equal than the timestamp of $d$.

UST achieves the same dependency meta-data overhead for remote updates as GentleRain, but it represents  a considerable improvement over Cure, which needs to store a vector of size equal to the number of data centers with each remote data item. }

{\color{black}
Unlike GentleRain, instead, UST requires that the local copy of an item $d$ stores a dependency vector with one entry per data center (with the local entry corresponding to the timestamp of $d$). 
 By this vector, \ts{} can determine, at the data center granularity, the snapshot of the data store to which an item belongs. This allows \ts{} to implement RO-TX efficiently by overcoming a key limitation of the design of GentleRain, which only stores the timestamps of local items.

\pvs
~\\\noindent{\bf Support for fast RO-TX.} In \ts{}, when receiving a RO-TX request from a client, the transaction coordinator determines the snapshot vector of the transaction. 
 Such vector has one entry per data center and represents the freshest snapshot corresponding to stable items and including all the dependencies of the client. Every item belonging to such snapshot must be visible by the transaction. 
To determine whether a local item is visible to the transaction, \ts{} exploits the available item's dependency vector, and checks whether it is entry-wise smaller than or equal to the snapshot vector. 
To determine whether a remote item $d$ created at the $i$-th data center is visible to the transaction, \ts{} checks if the item timestamp is lower than or equal to the $i$-th entry in the transaction vector. This condition is sufficient because the transaction vector includes only stable items by construction. Hence, if the condition is met, UST ensures that all of $d$'s dependencies have already been received in the data center and are, hence, visible to the transaction.

In GentleRain, instead, the snapshot visible to a transaction is determined by a single snapshot timestamp. Every item with a timestamp lower than such value is visible to a transaction. A transaction's timestamp has to be higher than the highest dependency timestamp at the client to include all the client's dependencies. Let $t$ be the snapshot timestamp of a transaction. To enforce that the transaction does not ``miss" any item that it should be able to access, GentleRain must ensure that the local data center has received {\em all} items from {\em all} data centers with a timestamp lower than or equal to $t$. 
The duration of this synchronization step is potentially proportional to the communication delay between the local data center and the furthest data center.
}

\pvs
~\\\noindent{\bf Visibility latency.} The inevitable price to be paid for the increase in availability is that the visibility latency of remote updates is increased, because there needs to be communication between replicas in different data centers to compute visibility. We believe the availability gains well warrant the slight increase in update latency. 

\begin{table}[t!]
\begin{center}
{\scriptsize
\begin{tabular}{| >{\centering\arraybackslash}m{1.3cm} | >{\centering\arraybackslash}m{4.5cm}|}
\hline
{\bf Symbol}  & {\bf Definition} \\
\hline
N& \# partitions\\
\hline
M& \# replicas per partition\\
\hline
$p_n^m$&The $m-$th replica of the $n-$th partition \\
\hline
$dt_c$&Dependency time at client c\\
\hline
$GSV_n^m$ & Global stable vector on $p_n^m$ \\
\hline
$USV_n^m$ &  Universal stable vector on $p_n^m$\\
\hline
$USV_c$ &   Universal stable vector at client $c$\\
\hline
$Clock_n^m$& Physical clock time on $p_n^m$\\
\hline
$VV_n^m$& Hybrid version vector of $p_n^m$ \\ 
\hline
$d$&A tuple $\langle k, v, ut, sr, DV\rangle$\\
\hline
\end{tabular}
}
\caption{Definition of symbols.}
\label{tab:notations}
\end{center}
\end{table}

\section{Protocols in Okapi}
\label{sec:protocol}
We now describe in detail the protocols run by \ts{}\footnote{The correctness proof is omitted for space constraint.}. Algorithm~\ref{alg:clnt} and  Algorithm~\ref{alg:srv} describe, respectively, how clients and servers implement PUT, GET and RO-TX  operations. Algorithm~\ref{alg:hyb} describes the management of clocks on servers. Algorithm~\ref{alg:ust} reports the UST stabilization protocol.
We indicate a target client as $c$. At the beginning of a session, $c$ is provided with the id $m$ of the data center it is connected to, referred to as the local data center. We refer to the server that handles $c$'s request as $p_n^m$. $p_n^m$ can be the node with which $c$ has established a session, or the node to which the request has been forwarded (as described in Section~\ref{sec:def}).  Table~\ref{tab:notations} provides a summary of the symbols used in the discussion.
\subsection{Meta-data}
\label{sec:protocol:metadata}
{\bf Item.}  An item $d$ is a tuple $\langle k, v, ut, sr, DV\rangle$. $k$ is the unique id that identifies the key of which $d$ is a version. $v$ is the value of $d$. $sr$ is the {\em source replica} of $d$, i.e., the id of the data center in which $d$ has been created. $ut$ is the update time, i.e., the creation time of the $d$ at its source replica. $DV$ is a dependency vector with $M$ entries. For a local update, $DV[i]$, $i\neq m$, is the update time of the item $d'$ with the highest timestamp such that $i)$ $d'$ has originated at the $i-$th replica and $ii)$ $d$ depends on $d'$. $DV[m]$ is equal to $ut$. For remote updates, $DV$ is null.

\begin{algorithm}[t!]
\algsize
\caption{\ts{} client $c$ at data center $m$.}
\label{alg:clnt}
\begin{algorithmic}[1]

\Function{GET}{key $k$}\label{alg:clnt:get}
	\State send \param{GETReq \mib{k, USV_c}} to server
	\State receive \param{GETReply \mib{v, USV_n^m, ut, sr}}
	\State $USV_c \lam max\{USV_n^m, USV_c\}$\label{alg:clnt:get:usv}
	\State {\bf if} ($sr == m$) {\bf then} $dt_c = max\{dt_c,\ ut\}$ {\bf endif}\label{alg:clnt:get:dtc}
	\State \Return v
\EndFunction

\Statex 

\Function{PUT}{key $k$, value $v$}\label{alg:clnt:put}
	\State $DV_c \lam USV_c$; $DV_c[m] \lam max\{dt_c, DV_c[m]\}$
	\State send \param{PUTReq \mib{k,v, DV_c}} to server\label{alg:gr:client:put:send_DV}
	\State receive \param{PUTReply \mib{ut}}
	\State $dt_c\lam ut$\Comment{Update client's dependency at local data center}\label{alg:clnt:put:dtc}
\EndFunction

\Statex
\Function{RO-TX}{key-set $\chi$}\label{alg:clnt:tx}
	\LineComment{Send remote dependencies ($USV_n^m$) and local dependencies ($dt_c$) info}
	\State send \param{RO-TX-Req \mib{\chi, USV_c, dt_c}} to server \mib{p_m^n}
	\State receive \param{RO-TX-Resp \mib{D, USV_n^m}}
	\State $USV_c \lam max\{USV_n^m, USV_c\}$\label{alg:clnt:get:usv}
	\For{($d \in D$)}	
		\State read $d$ as if it were the result of a GET\Comment{This updates $dt_c$ if necessary}
	\EndFor
\EndFunction
\end{algorithmic}
\end{algorithm}

\begin{algorithm}[t]
\algsize
\caption{\ts{} server $p^m_n$ serving clients requests.}
\label{alg:srv}
\begin{algorithmic}[1]
\Event{receive \param{GETReq \mib{k, USV_c}} from $c$}\label{alg:srv:get}
		\State $USV_n^m \lam max\{USV_n^m, USV_c\}$\label{alg:srv:get:usv}
		\State $D_k \lam \{ d : d.k == k \}$\Comment{Versions chain of the desired  key}
		\LineComment{Visible version with highest timestamp}
		\State $d \lam argmax_{d.ut}\{D\} : (d.sr == m \ \lor \ d.ut \leq USV_n^m[d.sr])$\label{alg:srv:get:get}
	\State send \param{GETReply \mib{USV_n^m,d.v,d.ut,d.sr}} to client
\EndEvent

\Statex

\Event{receive \param{PUTReq \mib{k,v,DV_c}} from $c$}\label{alg:srv:put}

\State $updateClockOnPut(DV_c)$ \Comment{Update version vector}\label{alg:srv:put:clock}
\State  $d.k \lam k;\ d.v\lam v;\ d.ut \lam VV_n^m[m];\ d.sr \lam m;\ d.DV \lam DV_c$
\State  $d.DV[m]\lam d.ut$
\State  insert $d$ in the version chain of key $k$
\State  send \param{PUTReply \mib{d.ut}} to client
\For{($i \lam 0\ldots M, i\neq m$)}
\State send \param{\mib{Replicate} d.k, d.v, d.ut} to $p_n^i$
\EndFor
\EndEvent
\State lastOutMsg $\lam Clock_n^m$
\Statex

\Event{receive \param{RO-TXReq\ \mib{\chi, USV_c, dt_c}} from $c$} \label{alg:srv:xact}
\State $updateClockOnTx(dt_c)$ 
\State $USV_n^m\lam max\{USV_c, USV_n^m\}$\Comment{Install newer USV if needed}
\State $lts\lam VV_n^m[m]$ \Comment{Take freshest local snapshot}
\State $\chi_i \lam \{k \in \chi : partition(k) == i\}$\Comment{Set of requested keys per node}
\State $D \lam \emptyset$ \Comment{Items to return to client}
\For{($i\ s.t.\ \chi_i\neq \emptyset$)}\Comment{Done in parallel}
	\State send \param{\mib{SliceREQ\ \chi_i, lts, USV_n^m}} to $p_i^m$ 
	\State receive \param{\mib{SliceRESP\ D_i}} from $p_i^m$
	\State $D \lam D \cup D_i$
\EndFor
\State reply \param{\mib{D, USV_n^m}} to $c$
\EndEvent

\Statex

\Event{receive \param{SliceREQ\ \mib{\chi, lts, USV_i^m}} from the coordinator $p_i^m$}\label{alg:srv:slice}
\State $updateClockOnTx(lts)$ \Comment{Update $Clock_n^m$ to cope with clock skew. }
\State $USV_n^m\lam max\{USV_i^m, USV_n^m\}$\Comment{Install newer USV if needed}
\State $TS \lam USV_i^m; TS[m]\lam lts$\Comment{Transaction snapshot vector}\label{alg:srv:slice:ts}
\State $D \lam \emptyset$
\For{$k \in \chi$}
\State $D_k \lam \{ d: d.k == k \land \big( (d.sr == m \land d.DV\leq TS) \ \lor$\label{alg:srv:slice:loc}
\State $ \ \ \ \ \ \ \ \ \ \ \ \  \ \ \ \ \ \ \ \ \ \  \ \ \ \ \ \ \ \ \ \ \ \ \ \ \ \ \ \ \ \ \ \ \ \ \ \  (d.sr\neq m \land d.ut \leq TS[d.sr])  \big)\}$\label{alg:srv:slice:rem}
\State $D \lam D \cup  argmax_{d.ut} \{D_k\}$\Comment{Freshest visible version}
\EndFor
\State reply \param{\mib{SliceRESP\ D}} to $p_i^m$
\EndEvent

\Statex

\Event{receive \param{Replicate\ \mib{k, v, ut}} from $p_n^i$}\label{alg:repl}
\State create new item $d$
\State $d.k\lam k$; $d.v\lam v$; $d.ut\lam ut$; $d.sr\lam i$
\State insert $d$ in the version chain of key $d.k$
\State $VV_n^m[i]\lam d.ut$
\EndEvent
\end{algorithmic}
\end{algorithm}

\pvs
~\\\noindent{\bf Client.} A client $c$ maintains one vector $USV_c$ with one entry per data center. $USV_c$ indicates the freshest stable snapshot from which any server has served a GET or a read-only transaction issued by $c$. $c$ also maintains a {\em dependency time} $dt_c$, corresponding to the highest timestamp of any local item read or written by $c$.

\pvs
~\\\noindent{\bf Server.}  A server $p_n^m$ has access to a monotonically increasing physical clock, $Clock_n^m$. 
 $p_n^m$ also maintains three vector clocks with $M$ entries: $VV_n^m$, $GSV_n^m$ and $USV_n^m$.
$VV_n^m$ is a version vector of hybrid clocks. $VV_n^m[i], i\neq m$, indicates the timestamp of the latest update/heartbeat received by $p_n^m$ that comes from the replica at the $i-$th data center. $VV_n^m[m]$ is the version clock of $p_n^m$ and it is used to timestamp updates. 
$GSV_n^m[i] = t$ means that $p_n^m$ is aware that all the nodes in the $m-$th data center have processed all events generated in the $i-$th data center with timestamp up to $t$.
$USV_n^m[i] = t$ indicates that $p_n^m$ is aware that every node in every data center has installed all the updates generated in data center $i$ whose timestamps are smaller than or equal to $t$. {\color{black}{$GSV_n^m$ and $USV_n^m$ are read and written atomically. \ts{} uses optimistic locking}~\cite{Lim:2014} to keep the corresponding overhead low.}

\subsection{Operations}
\label{sec:protocol:operations}
{\bf GET.} $c$ sends a request $\langle$GET $k$, $USV_c\rangle$, where $k$ is the key to be read. $p_n^m$ uses $USV_c$ to advance $USV_n^m$ if necessary,  so as to be sure to install a snapshot that is at least as fresh as the one that $c$ has been served from so far.
$p_n^m$ then selects the version $d$ of $k$ with the highest timestamp such that either $d$ is local or $d$'s update time is smaller than or equal to the entry in $USV_n^m$ corresponding to $d$'s originating data center.
$p_n^m$ returns $USV_n^m$ and $d$'s value, timestamp and source replica to $c$.  Upon receiving such reply, $c$ updates $USV_c$ and, if $d$ is local, $dt_c$.

\pvs
~\\\noindent{\bf PUT.} $c$ sends a request $\langle$PUT $k,v,DV_c\rangle$, where $k$ is the key to be written and $v$ is the desired value to associate with $k$. $DV_c$ is a dependency vector whose remote entries are equal to the ones in  $USV_c$; the local entry is the maximum between the local entry in $USV_c$ and $dt_c$. $DV_c$ represents all  dependencies established by $c$ so far.

Upon receiving $c$'s request, $p_n^m$ first determines the hybrid timestamp to associate with the new update. To this end, $p_n^m$ invokes the \texttt{updateClockOnPut} function (reported in Algorithm~\ref{alg:hyb}). This function advances the local entry of the version clock of $p_n^m$, $VV_n^m[m]$, with a hybrid timestamp that is higher than the highest entry in $DV_c$ and than the current version clock $VV_n^m[m]$.
Then, $p_n^m$ creates a new version $d$ of $k$, and replies to $c$ with $d$'s timestamp. This is used by $c$ to update $dt_c$. Finally, $p_n^m$ replicates $d$ by sending to its replicas a copy of $d$, except $d.DV$.

Upon receiving such replication message, a replica $p_n^i$ inserts a copy of $d$ in the version chain corresponding to $d.k$ and sets $VV_n^i[m] = d.ut$.

\begin{algorithm}[t]
\algsize
\caption{\ts{} server $p^m_n$: clock management.}
\label{alg:hyb}
\begin{algorithmic}[1]
\Function{updateClockOnPut }{$DV_c$}\label{alg:hyb:new}
\State $hd \lam max\{ DV_c\}$ \Comment{Find highest dependency}
\State $max_p\lam max\{VV_n^m[m].p, Clock_n^m, hd.p\}$\Comment{Max physical clock}
\If{($max_p == VV_n^m[m].p == DV_c[m].p$)}\Comment{Local phys clock behind}
\State $l = max\{VV_n^m[m].l, DV_c[m].l\}+1$\\
\ \ \ \ \ \ {\bf else if} ($max_p == VV_n^m[m].p$) {\bf then} $l = VV_n^m[m].l + 1$\\
\ \ \ \ \ \ {\bf else if} ($max_p == DV_c[m].p$) {\bf then} $l = DV_c[m].l + 1$\\
\ \ \ \ \ \ {\bf else} $l = 0$\Comment{Local phys clock higher than dependency}
\EndIf
\State $VV[m].p \lam max_p; VV[m].l \lam l$
\EndFunction

\Statex

\Function{updateClockOnTx}{$ts$}\label{alg:hyb:update}
\State {\bf if} ($ts > VV[m] \land ts > Clock_n^m$) \ {\bf then} \  $VV[m] \lam ts$ {\bf endif}
\EndFunction

\Statex

\Function{updateClockOnHeartbeat}{}\label{alg:hyb:heart}
\If{($Clock_n^m > VV_n^m[m].p$)}
\State $VV_n^m[m].p \lam Clock_n^m; \ VV_n^m[m].l \lam 0$
\EndIf
\EndFunction

\Statex

\Event{every $\Delta$ time}\label{alg:heart:send}
\If{$Clock_n^m$ \mib{\geq lastOutMsg}+$\Delta$}
\State $updateClockOnHeartbeat()$
	\For{each server \mib{p_n^j}, $j\in \{0\ldots M-1\}, k\neq m$}
		\State send \param{HEARTBEAT \mib{VV_n^m[m]}} to \mib{p_n^j}
	\EndFor
\EndIf
\State lastOutMsg $\lam Clock_n^m$
\EndEvent

\Statex

\Event{receiving \param{HEARTBEAT ct} from \mib{p_n^j}}\label{alg:heart:recv}
\State \mib{VV_n^m}[j]\la ct
\EndEvent
\end{algorithmic}
\end{algorithm}

\pvs
~\\\noindent{\bf RO-TX.}  $c$ sends a request $\langle RO-TXReq, \chi,$ $USV_c, dt_c\rangle$, where $\chi$ is the set of keys to be read.
$USV_c$ and $dt_c$ are provided so that the transaction is served from a snapshot that includes $c$'s dependencies.

Upon receiving $c$'s request, $p_n^m$ acts as the coordinator for the corresponding transaction. First, $p_n^m$ computes the local transaction snapshot time, $lts$. This time represents the highest timestamp of local items visible to the transaction. $lts$ is computed as the maximum between the local clock at the coordinator and $dt_c$. $p_n^m$ also updates $USV_n^m$ if necessary. In this way, the snapshot defined by $USV_n^m$ and $lts$ is the freshest snapshot that includes all the dependencies established by $c$. $p_n^m$ sends $USV_n^m$ and $lts$ to every node $p_i^m$ that holds at least one key in $\chi$, together with the set of keys to be read.

Upon receiving such message, $p_i^m$ invokes the \texttt{updateClockOnTx} function (reported in Algorithm~\ref{alg:hyb}). This function advances $VV_i^m[m]$ in case it is lower than $lts$. $p_i^m$ also updates its $USV_i^m$ if it is smaller than the one proposed by the coordinator. Then, $p_i^m$ computes the transaction's snapshot vector $TV$ starting from the $USV$ and $lts$ proposed by $p_n^m$. $TV$ is equal to $USV_n^m$ in the remote entries. The local entry is, instead, the local transaction timestamp proposed by $p_n^m$. 
For each key to be read, $p_i^m$ determines the version $d$ with the highest timestamp such that $d$ is visible to $c$ according to $TV$.

\begin{algorithm}[t!]
\algsize
\caption{\ts{} server $p^m_n$: GSV and USV computation.}
\label{alg:ust}
\begin{algorithmic}[1]
\Event{every $\Delta_G$ time}\label{alg:gsv}
\State $GSV_n^m[j] \lam min\{VV_i^m[j]\}, \forall j=0\ldots M-1, \forall i = 0\ldots N-1$
\EndEvent

\Statex

\Event{every $\Delta_U$ time}\label{alg:usv}
\State $V[j] \lam min\{GSV_n^i[j]\}, \forall j=0\ldots M-1, \forall i = 0\ldots N-1$\label{alg:usv:min}
\State $USV_n^m \lam max\{V, USV_n^m\}$\label{alg:usv:mono}\Comment{Enforce monotonicity of USV}
\EndEvent
\end{algorithmic}
\end{algorithm}

A local item is visible if its update time and its dependencies fall within the boundaries defined by $TV$ (Line~\ref{alg:srv:slice:loc}). A remote item is visible if its update time falls within the boundaries of $TV$  (Line~\ref{alg:srv:slice:rem}). Since $TV$ is computed starting from a $USV$, this condition implies that the remote item is also stable. 

The set of all read items is sent back to $p_n^m$. Upon collecting all such replies, $p_n^m$ forwards them back to $c$ together with $USV_n^m$. Finally, $c$ updates $USV_n^m$ and, for each item in the returned set, updates its dependency meta-data as when processing the result of a GET.

\pvs
~\\\noindent{\bf Heartbeats.}  If $p_n^m$ does not receive update requests from clients, it does not send replication messages to its replicas either. Therefore, other replicas cannot increase the $m$-th entry in their version vector, and the $m-$th entry of the $USV$ cannot advance. 
To avoid this scenario, a partition that does not receive updates for a period of time longer than $\Delta$, broadcasts its latest local hybrid version clock time to its replicas.
The function \texttt{UpdateClockOnHeartbeat} computes the heartbeat timestamps by advancing the local version clock $VV_n^m[m]$ to $Clock_n^m$ if $Clock_n^m$ is higher than $VV_n^m[m]$. 
Heartbeat messages and update replication messages are sent (and received) in order of increasing update timestamps and clock values.
Upon receiving a heartbeat with timestamp $t$ from $p_n^m$, $p_n^i$ sets $VV_n^i[m] = t$. 

\begin{figure*}[t !]
 \subfloat[Throughput scalability.\label{fig:scal:xput}]{
\includegraphics[scale=0.5]{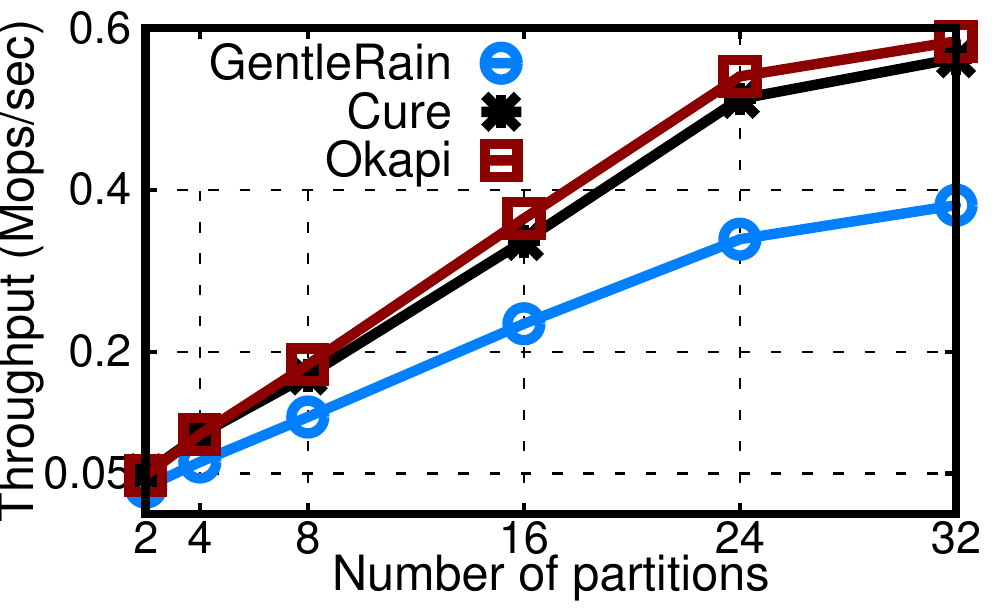}
}
\hfill
\subfloat[RO-TX avg. resp. time (log).\label{fig:scal:tx}]{
\includegraphics[scale=0.5]{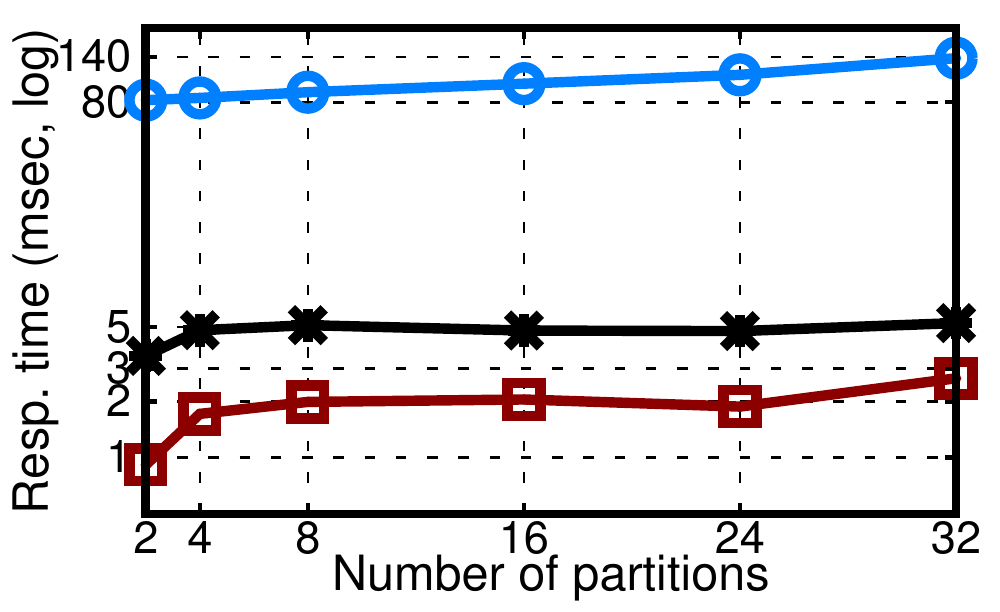}
}
\hfill
\subfloat[PUT avg. resp. time (log). \label{fig:scal:put}]{
\includegraphics[scale=0.5 ]{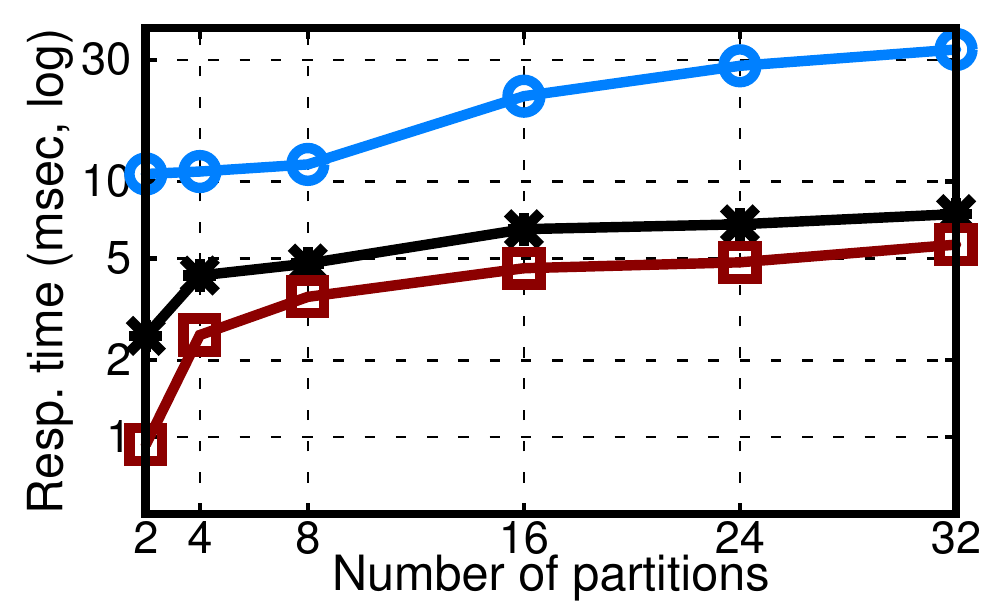}
}
\caption{Performance with increasing scales of the system. Clients perform a RO-TX involving two partitions and a PUT on a random partition. \ts{} achieves better or similar peak throughput with respect to GentleRain and Cure but considerably lower latencies. This is because, thanks to HLC, \ts{} never blocks  when serving an operation.}
\label{fig:scal}
\end{figure*}

\pvs
~\\\noindent{\bf Stabilization protocol.} Every $\Delta_G$ time units, partitions within a data center exchange their version vectors. $p_n^m$ computes $GSV_n^m$ as the aggregate minimum of known version vectors. Similarly to previous work~\cite{Du:2014,Akkoorath:2016}, \ts{} organizes nodes within a data center as a tree to reduce message exchange. 
 Every $\Delta_U$ time units, replicas at different partitions exchange their $GSV$ and compute the $USV$ as the aggregate minimum of the received $GSV$. Because $USV_n^m$ is also updated when serving client requests, it can happen that $USV_n^m$ becomes greater than $GSV_n^m$ in some entries. Thus, $p_n^m$ enforces that entries in $USV_n^m$ are monotonically increasing.

\pvs
~\\\noindent{\bf Garbage collection.} 
Servers within a data center periodically exchange the  transaction snapshot vector corresponding to their oldest active transactions and compute the garbage collection vector $GV$ as the entry-wise minimum of those vectors. If no transaction is active on $p_n^m$, $p_n^m$ sends a fake transaction snapshot vector, as computed in Algorithm~\ref{alg:srv} Line~\ref{alg:srv:slice:ts}. 
 A server retains every version of any key $k$ it stores up to and including the freshest version that would be visible to a transaction with transaction vector $GV$.  
  Older versions are removed. That is, \ts{} retains up to the oldest version of $k$ that could still potentially be visible to a transaction.

\section{Evaluation}
\label{sec:eval}
\subsection{Methodology and performance metrics}
{\color{black}
We evaluate \ts{} by responding to these questions:
\begin{itemize}
\setlength\itemsep{0pt}
\item How well does \ts{} scale?
\item What  throughput  can  \ts{} achieve?
\item What latencies do \ts{} operations achieve?
\item How much does \ts{} benefit from HLC?
\item What is the penalty in update visibility latency incurred by \ts{} to support higher availability?
\item What is the communication overhead of UST?
\end{itemize}

We answer these questions by comparing \ts{} with Cure and GentleRain on a large scale public cloud infrastructure. We evaluate the performance of  these systems with different workloads and deployment settings. We report  achievable throughput and average operation latencies, with a focus on PUT and RO-TX operations, since GET operations are not affected by clock skew. We also report remote updates visibility latency, communication costs and dependency tracking meta-data overhead.} {\color{black}We conduct our evaluation using a benchmark that allows us to accurately assess the sensitivity of \ts{} to key workload characteristics like the number of partitions involved in a transaction and write intensity.}

\subsection{Experimental test-bed}
\label{sec:eval:test-bed}
We consider an Amazon AWS deployment with 3 data centers and 32 partitions each. The data centers are in Oregon, N.Virginia and Ireland. 
We use $c4.large$ instances (2 virtual CPUs and 3.75 GB of RAM). Data is stored in-memory, without any fault tolerance mechanism. This allows us to evaluate \ts{} without taking into account the dynamics and overhead of logging/replication. \ts{} can be extended to achieve fault tolerance by means of standard techniques~\cite{Oki:1988,Lamport:1998,vanRenesse:2004}.

\begin{figure*}[t!]
\subfloat[Throughput (log).\label{fig:xact:xput}]{
\includegraphics[scale=0.4]{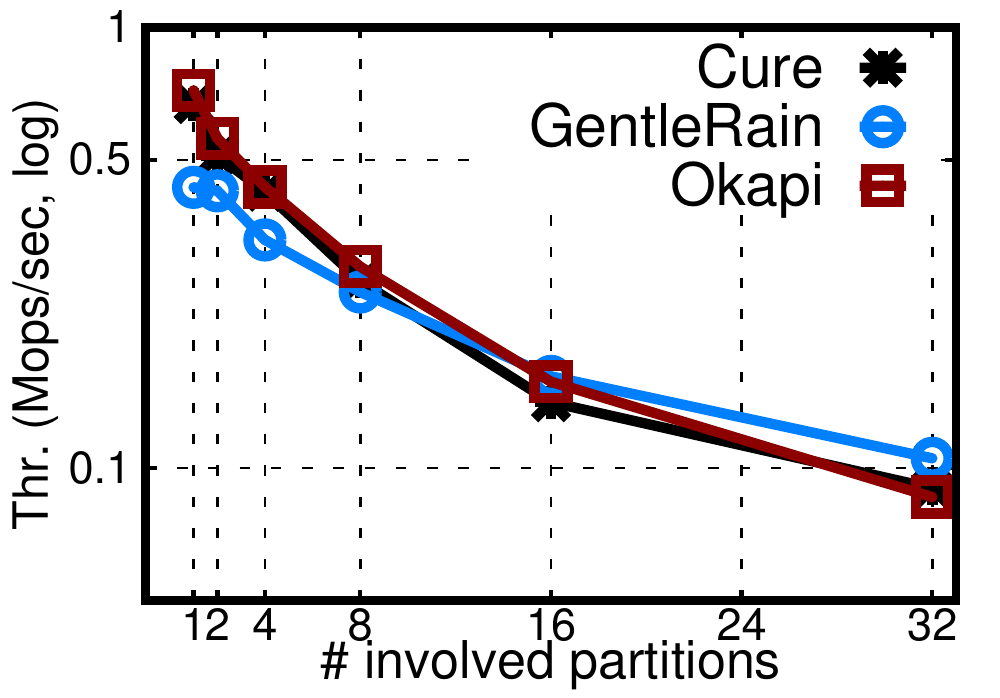}
}
\hspace{-0.45cm}
\subfloat[RO-TX avg. resp. time (log).\label{fig:xact:resp}]{
\includegraphics[scale=0.4]{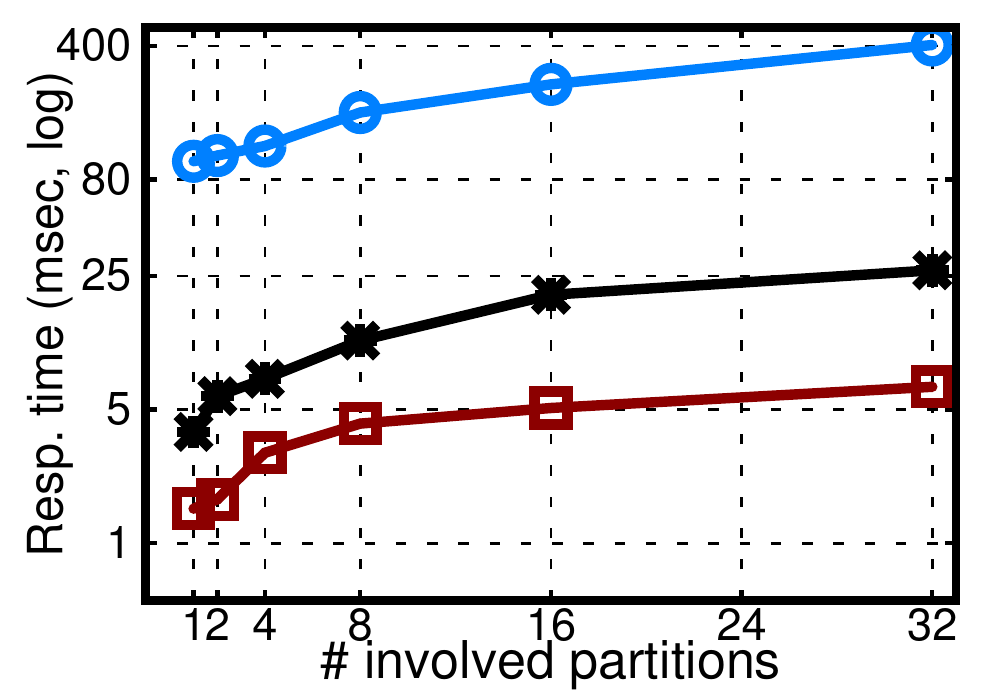}
}
\hspace{-0.45cm}
\subfloat[RO-TX wait probability.\label{fig:xact:block_prob}]{
\includegraphics[scale=0.4]{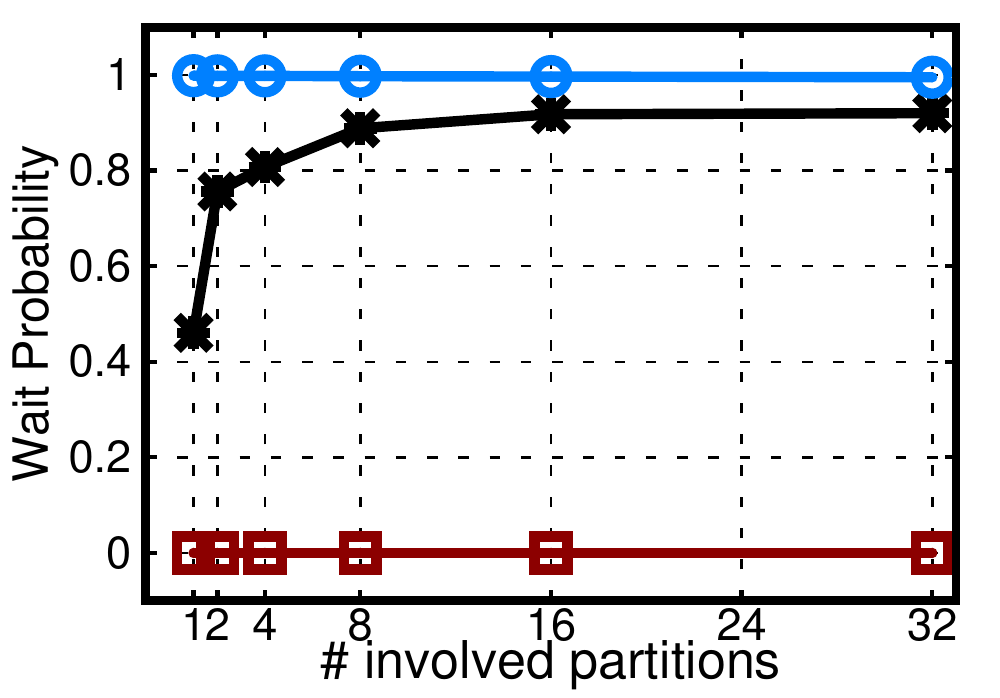}
}
\hspace{-0.45cm}
\subfloat[RO-TX avg. wait time (log).\label{fig:xact:block_time}]{
\includegraphics[scale=0.4]{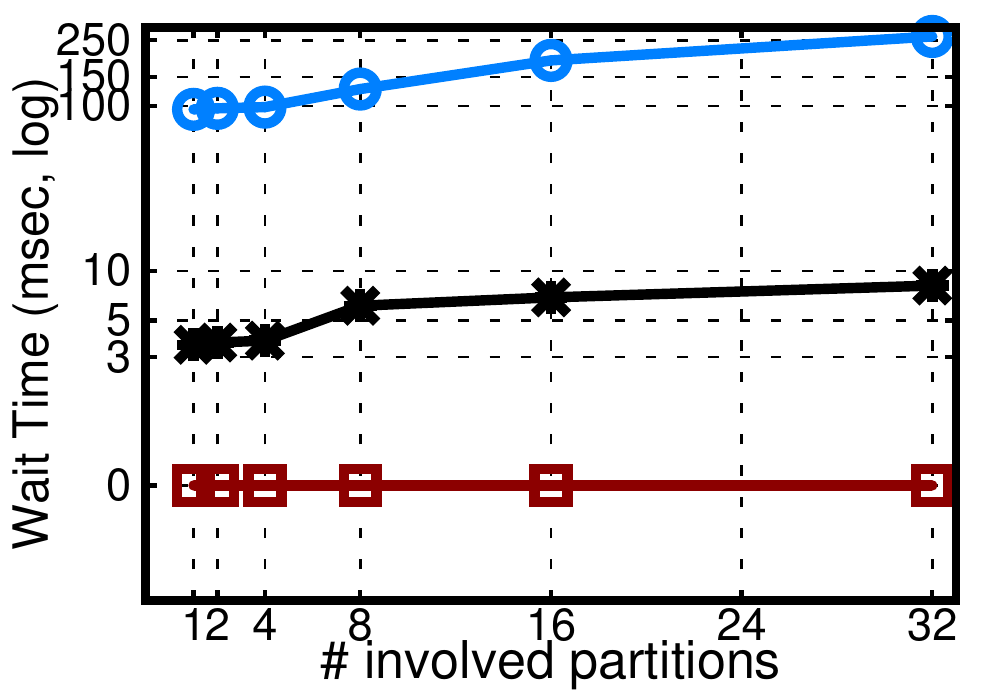}
}
\caption{Performance while varying the number of partitions involved in a transaction. Clients perform a RO-TX and a PUT touching random partitions. \ts{} achieves better or comparable peak throughput with respect to GentleRain and Cure, but achieves considerably lower latencies. By means of HLC and UST, \ts{} never blocks a transaction. Instead, Cure stalls transactions because of clock skew among nodes in the {\em local} data center. GentleRain incurs the highest waiting time and probability because it has to wait to receive {\em all} the items from {\em all the remote} data centers that are included in the snapshot visible to a transaction.}
\label{fig:xact}
\end{figure*}

Each partition is composed of one million key-value pairs. We consider small items, with keys and values of 8 bytes, as representative of many production workloads~\cite{twitter,instagram,Atikoglu:2012,Nishtala:2013}. Keys are chosen within each partition according to a zipf distribution with parameter 0.99. Clients are collocated with servers, establish their sessions with the collocated server and perform operations in closed loop.  
 We run NTP~\cite{ntp} to synchronize  physical clocks. 
 As in previous work~\cite{Akkoorath:2016}, clocks are synchronized before each experiment. We use the NTP server 0.amazon.pool.ntp.org.
All the stabilization protocols are run every 5 milliseconds. Heartbeats are sent by a node if it does not serve any put request for 1 millisecond. 

{\color{black}Hybrid timestamps, similarly to physical ones, are encoded with 64 bits. If a node has to increase the logical part of its HLC but $HLC.l$ has already reached the maximum value, the node has to resort to waiting. We use the 48 most significant bits of a HLC as physical part and the other 16 as logical part. As such, a hybrid timestamp can track phyisical time up to microsecond granularity and can encode up to $2^{16}$ logical events~\cite{Kulkarni:2014}.  With this settings, we have never witnessed a node resort to waiting.}

\subsection{Experimental results}
\label{sec:eval:results}
{\bf Scalability.} We evaluate the scalability of \ts{} by running a workload on an increasing number of partitions, from 2 to 32. In this workload, each client performs a RO-TX involving two partitions (so as to keep the number of contacted partitions fixed regardless of the scale) and a PUT. The partitions touched by the operations are chosen uniformly at random. Figure~\ref{fig:scal} depicts the result of the experiment, reporting peak throughput in Figure~\ref{fig:scal:xput}, average response time of the RO-TX operation in Figure~\ref{fig:scal:tx} and of the PUT operation in Figure~\ref{fig:scal:put}.

\ts{} achieves 50\% higher throughput than GentleRain and a slightly higher throughput than Cure. \ts{}, however, achieves much lower latencies for both RO-TX and PUT operations, up to two orders of magnitude lower than GentleRain and 100\% lower than Cure.

\ts{} achieves this result by never blocking operations. Cure and GentleRain need to activate many more clients than \ts{} to compensate for the idle waiting times and saturate their resources.  
 \ts{} and Cure use vector clocks, so their peak throughput is similar. \ts{}'s throughput is slightly higher because UST allows for better resource efficiency. The use of scalar clocks leads GentleRain to incur very long waiting times to serve a transaction, as explained in Section~\ref{sec:design:dep}. The excessive number of client threads, needed to fill the long waiting times, limits GentleRain's overall scalability.
 
\pvs
~\\\noindent{\bf Sensitivity to RO-TX characteristics.} We now evaluate the performance of \ts{} when serving transactions that span different numbers of partitions. To this end, we consider a workload in which clients issue a RO-TX to read $p$ keys and then write one key belonging to a random partition. Each read key is stored on a different partition, and partitions are chosen uniformly at random. We fix the number of partitions per data center to 32 and we analyze the performance of the three systems while varying $p$ from 1 to 32.

\begin{figure}[b!]
\centering
\includegraphics[scale=0.7]{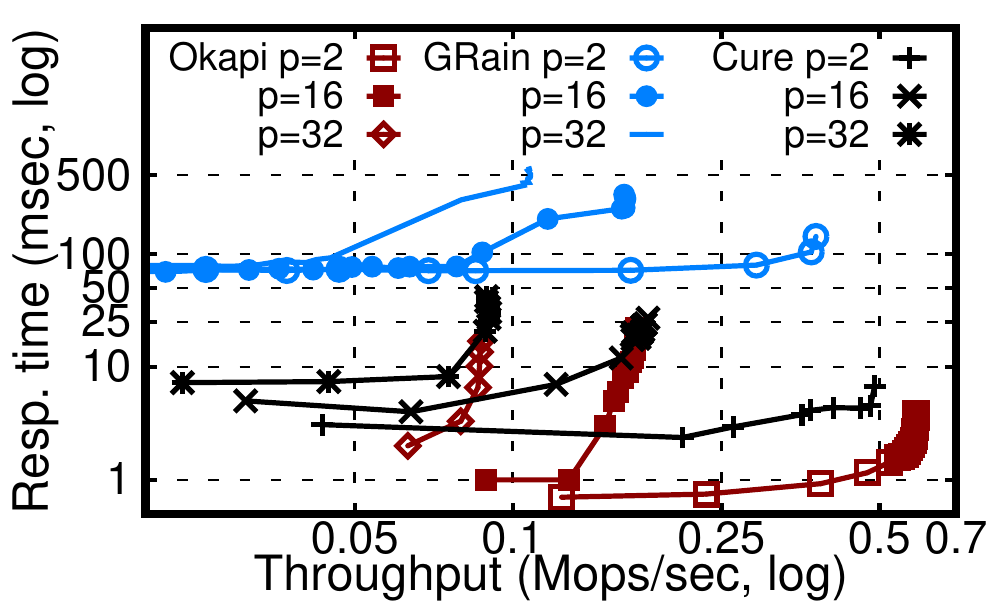}
\caption{RO-TX avg. resp. time (log) as a function of the throughput and of the \# partitions involved in a transaction ($p$). \ts{} achieves the lowest latency and almost always the highest throughput. For $p = 32$ \ts{} attains a slightly lower throughput than GentleRain, but achieves a 2 orders of magnitude lower latency.}
\label{fig:xact:vs}
\end{figure}

Figure~\ref{fig:xact:xput} shows the throughput achieved by the considered systems. Figure~\ref{fig:xact:resp} depicts the average transaction response time corresponding to the throughput values of Figure~\ref{fig:xact:xput}. Figure~\ref{fig:xact:block_prob} and Figure~\ref{fig:xact:block_time} report, respectively, the probability that a transaction is stalled before being served and the duration of the stall. Figure~\ref{fig:xact:vs} reports, for different values of $p$, the average RO-TX response time as a function of the throughput.

The plots show that \ts{} achieves a slightly better throughput than Cure, for any value of $p$. \ts{} is up to 60\% better than GentleRain for transactions that span up to 8 partitions. Then GentleRain achieves a marginally higher throughput than Okapi. Cure and GentleRain, however, incur considerably higher latencies because of their blocking behavior, for any value of throughput.
GentleRain achieves a higher throughput when the number of contacted partitions is high because it timestamps transactions with a scalar and not with a vector, as in \ts{} and Cure. This results into a lower utilization of the network, which enables more concurrency when a transaction involves many partitions.

The plots also show the different blocking behaviors of Cure and GentleRain. In Cure, the probability of waiting due to clock skew increases with the number of contacted partitions. 
 The waiting time is proportional to the clock skew, and it is in the order of 5-10 milliseconds on average. GentleRain always waits to receive from all the data centers all the items that are in the transaction snapshot. The waiting time is, hence, mainly proportional to the communication latency with the furthest data center.

\begin{figure}[t!]
\subfloat[Throughput.\label{fig:pg:put_xput}]{
\includegraphics[scale=0.42]{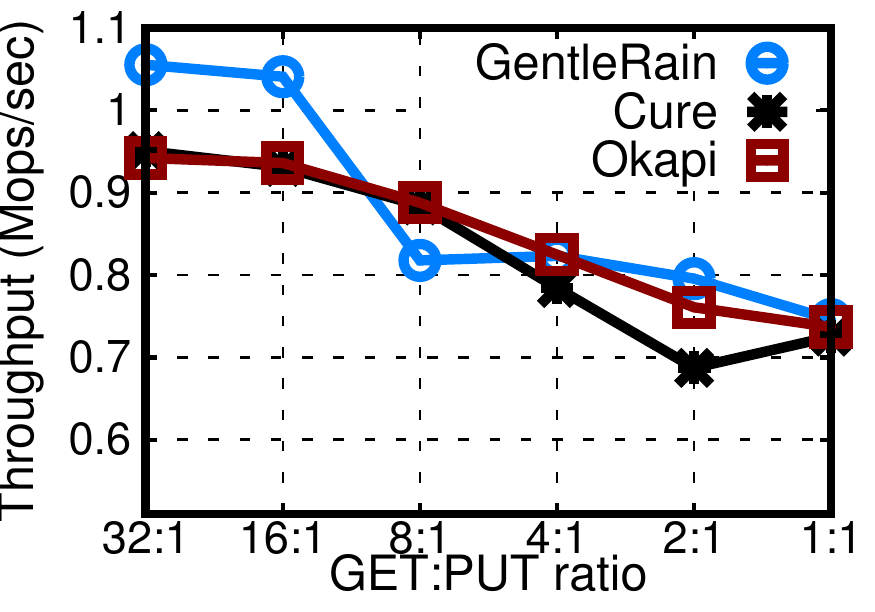}
}
\subfloat[PUT wait probability.\label{fig:pg:put_wait}]{
\includegraphics[scale=0.42]{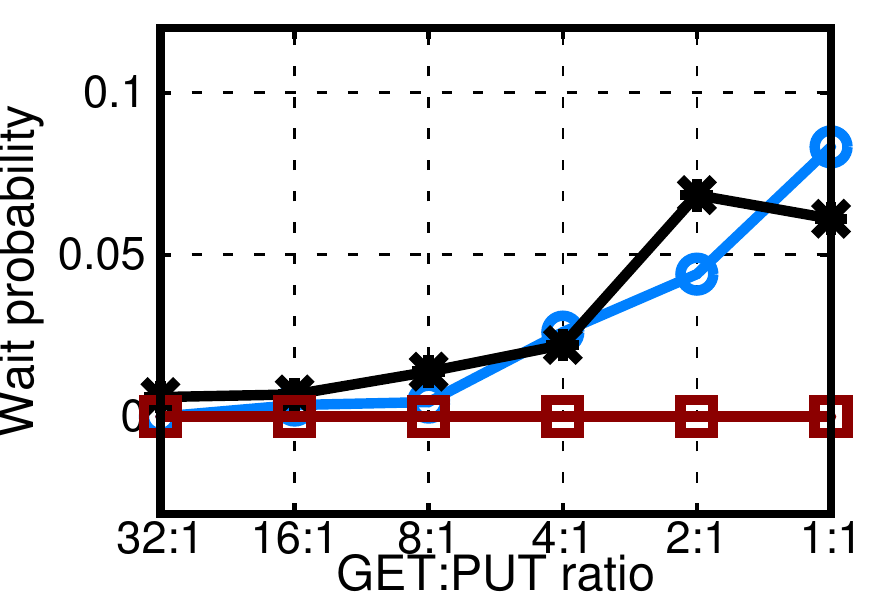}
}
\caption{Performance of transaction-less workloads with different GET:PUT ratios (32 partitions). \ts{} never blocks PUT operations and thus performs slightly better than Cure. GentleRain achieves slightly higher throughput in read-dominated scenarios because it only uses scalar dependency timestamps instead of vectors. \ts{} trades this marginal penalty for much bigger gains in RO-TX latencies and higher availability.}
\end{figure}

\pvs
~\\\noindent{\bf Sensitivity to write intensity.} We now analyze the sensitivity of the three systems to the workload write intensity. To this end, we run different workloads consisting of only GET and PUT operations, using 32 partitions. Clients read $g$ items on $g$ distinct partitions chosen uniformly at random and then update one item on a random partition. We vary $g$ from $1$ to $32$.  Figure~\ref{fig:pg:put_xput} reports peak throughput and Figure~\ref{fig:pg:put_wait} reports the probability that a PUT operation is stalled due to clock skew.

The plots show that \ts{} achieves a higher throughput than Cure. The difference between the two increases as the probability of stalling a PUT due to the clock skew increase with the write intensity of the workload.
\ts{} is, instead, comparable or competitive with GentleRain. In the most read-intensive workloads \ts{} incurs a slight throughput penalty ($<10\%$) because of the use of vector clocks instead of a single scalar. We believe this small cost is well worth the huge improvement that \ts{} attains in the RO-TX implementation and the higher level of availability that \ts{} achieves.

\begin{figure*}[t!]
\subfloat[Remote updates visibility latency.\label{fig:ust:vis}]{
\includegraphics[scale=0.52]{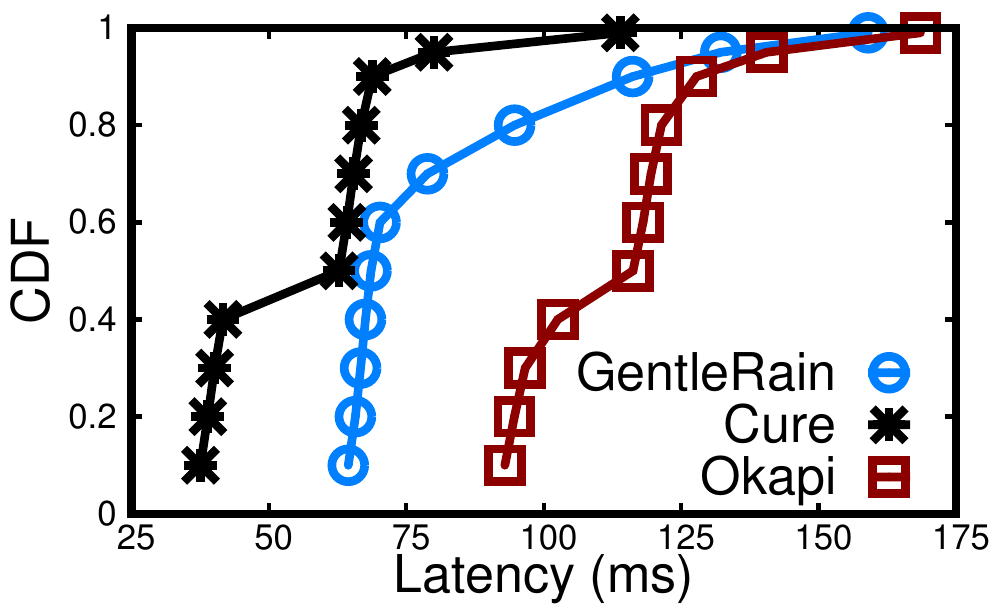}
}
\hfill
\subfloat[Bytes/replicated update.\label{fig:ust:ovhd}]{
\includegraphics[scale=0.52]{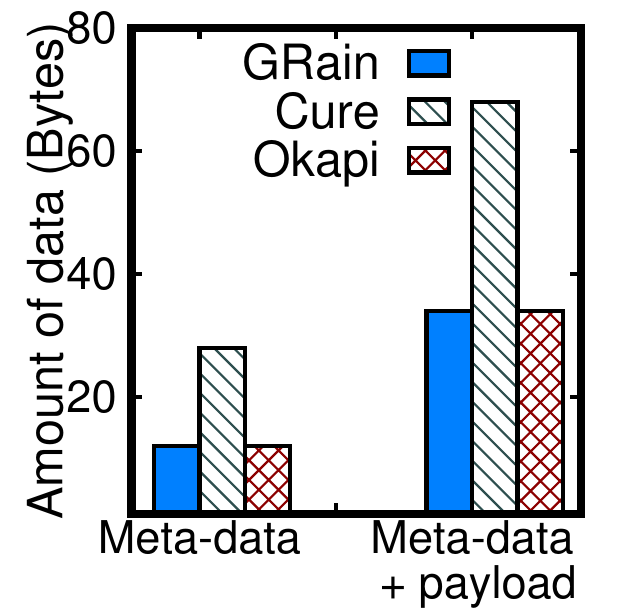}
}
\hfill
\subfloat[Data exchanged for the stabilization  protocol (log).\label{fig:ust:stab}]{
\includegraphics[scale=0.52]{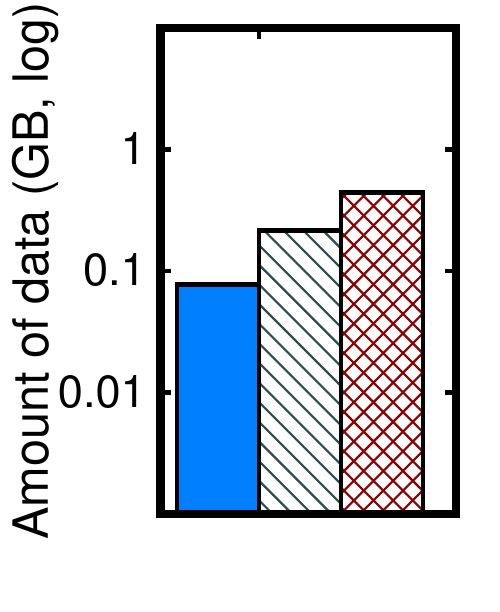}
}
\hfill
\subfloat[Data exchanged for the stabilization protocol and updates replication (log).\label{fig:ust:total}]{
\includegraphics[scale=0.52]{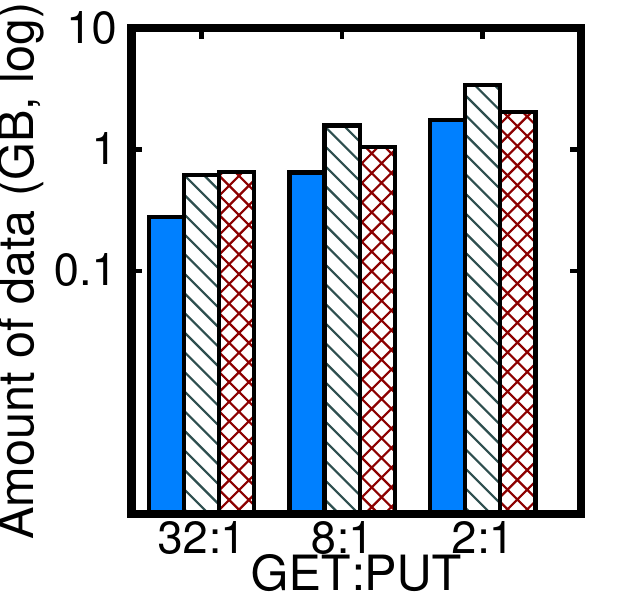}
}
\caption{Effects of UST. UST incurs a slightly higher visibility latency than Cure and GentleRain to support higher availability (a). As a by-product, UST matches the remote updates dependency tracking overhead of GentleRain, which only uses scalar clocks (b). \ts{}'s stabilization protocol exchange more data than Cure's and GentleRain's to achieve higher availability (c). This overhead is amortized by the reduction in meta-data for replicated updates (d).}
\end{figure*}

\pvs
~\\\noindent{\bf Implications of UST.} We now evaluate the effects of the stabilization protocols of the three systems.  Figure~\ref{fig:ust:vis} reports the CDF corresponding to the  visibility latencies of remote updates in the 32:1 GET:PUT workload on 2 partitions. \remove{\color{red}To generate the CDF e first compute on each node different percentiles of the visibility latency. Then, we compute the average of homologous percentile values.} Figure~\ref{fig:ust:ovhd} depicts the amount of data replicated per update. Figure~\ref{fig:ust:stab} reports the amount of data exchanged to execute the stabilization protocols and Figure~\ref{fig:ust:total} reports the total amount of data exchanged among nodes (for the stabilization protocol and updates replication) while varying the write intensity of the GET-PUT workload on 32 partitions. 

\ts{} achieves the highest remote update visibility latency, for the sake of higher availability, followed by Cure and GentleRain. In Cure, the visibility latency in data center $DC_R$ of an item $d$ originated in $DC_L$ depends on the delay between $DC_L$ and $DC_R$~\cite{Akkoorath:2016}. In GentleRain, instead, the visibility latency depends on the delay between $DC_R$ and its furthest data center~\cite{Du:2013}.
The CDF of \ts{} and Cure is bi-modal (one mode per remote data center) because the visibility latency of $d$ depends (also) on the communication latency between $DC_L$ and $DC_R$. GentleRain's CDF, conversely, is unimodal because the remote update visibility latency depends on the communication delay between $DC_R$ and its furthest data center.


Deferring the visibility of updates allows \ts{} to match the resource efficiency of GentleRain when replicating updates. \ts{} and GentleRain need only 12 bytes of meta-data, corresponding to the source replica (4 bytes) and update time (8 bytes). Cure needs additional 8 bytes for each remote entry of the dependency vector. In our setting, then, Cure requires 28 bytes of meta-data per update, which is more than two times the overhead incurred by \ts{} and GentleRain. 
In our experiments, an update contains additional 16 bytes to encode the key and the value. Even if the payload amortizes the meta-data overhead, the amount of data sent by Cure to replicate an update is still almost 60\% higher than in \ts{} and GenleRain. In \ts{} and GenleRain dependency meta-data for replicated updates is insensitive to the scale of the system. In Cure, instead, it grows linearly with the number of data centers in the system. 

~\\
UST requires an additional round of inter-data center communication to achieve higher availability. For this reason, the  stabilization protocol of Okapi is more expensive than Cure's and GentleRain's. Such overhead, however, is compensated for by the reduced meta-data overhead achieved by \ts{}. If we consider, in fact, the total amount of data exchanged, i.e., stabilization protocol overhead and replication cost, \ts{} incurs a  communication overhead similar to Cure in read dominated workloads, and lower than Cure as the write intensity increases. 
GentleRain's stabilization protocol is the most network efficient regardless of the write intensity of the workload. Its gains against UST, however, decrease as the write intensity increases and the dominant communication cost becomes the updates replication. 

\ts{} could significantly reduce the UST overhead by piggybacking the computation of the USV to the one of the GSV. We have not experimented with this design yet.
\section{Related Work}
\label{sec:rw}
Our work is primarily related to the literature on causally consistent systems.
The first breed of such systems includes Bayou~\cite{Petersen:1997}, lazy replication~\cite{Ladin:1992}, ISIS~\cite{Birman:1987}, causal memory~\cite{Ahamad:1995}, and PRACTI~\cite{Belaramani:2006}. They implement causal consistency but assume single-machine replicas and do not consider partitioned data-sets.
COPS~\cite{Lloyd:2011} represents the first in a new class of systems, which implement causal consistency for both replicated and partitioned data stores.  
This second set of systems includes Eiger~\cite{Lloyd:2013}, Bolt-on causal consistency~\cite{Bailis:2013}, ChainReaction~\cite{Almeida:2013}, Orbe~\cite{Du:2013}, GentleRain~\cite{Du:2014}, SwiftCloud~\cite{Zawirski:2015} and Cure~\cite{Akkoorath:2016}. 
\ts{} differs from these systems on two levels: event timestamping and  dependency tracking.

\pvs
~\\\noindent{\bf Event timestamping.} COPS, Eiger, ChainReaction, Bolt-on and Orbe  use logical clocks to timestamp items. These systems exchange explicit dependency check messages among partitions to verify that a remote update can be made locally visible. GentleRain and Cure, instead, use loosely synchronized physical clocks and implement a stabilization protocol to determine the visibility of remote updates. GentleRain and Cure achieve higher performance than previous systems but incur additional synchronization delays to cope with clock skew. By employing HLC~\cite{Kulkarni:2014}, \ts{} implements a cheap stabilization protocol and is insensitive to clock skew.
Concurrently to our work, the use of HLC to achieve causal consistency has also been investigated in GentleRain+~\cite{gr+}. GentleRain+ simply augments GentleRain with HLC to make PUT operations robust against clock skew. The stabilization protocol and the implementation of transactions are the same as in GentleRain, so GentleRain+ inherits the limitations of GentleRain that we have described in the paper. Conversely, \ts{} uses a novel combination of HLC and dependency vectors to implement efficient transactions.  As we have shown, this combination is paramount to achieve scalability and low-latency for production-like workloads, which rely on efficient snapshot reads. Moreover, \ts{} achieves higher availability than GentleRain+ thanks to UST.

\pvs
~\\\noindent{\bf Dependency tracking.} The systems based on logical clocks keep detailed dependency information, encoded as a dependency list~\cite{Lloyd:2011,Lloyd:2013,Almeida:2013,Bailis:2013} or matrix~\cite{Du:2013}.  {\color{black}The techniques proposed to reduce the resulting overhead have downsides like per-update acknowledgement messages among replicas~\cite{Du:2013}, call-backs to the client~\cite{Du:2013}, or delay the visibility of updates also in the local data center~\cite{Du:2014b,Zawirski:2015}.}
 GentleRain and Cure track dependencies at a coarser granularity. GentleRain uses a single timestamp to achieve minimal overhead but incurs high waiting times to serve read-only transactions. Cure uses dependency vectors to avoid this issue but incurs a dependency tracking overhead linear in the number of data centers. \ts{} uses dependency vectors too but reduces the meta-data for remote updates at the cost of slightly delaying their visibility at remote sites.

\vspace{-3pt}
~\\
\ts{}'s design is also related to the use of physical and hybrid clocks in systems that target different consistency guarantees, e.g., Spanner~\cite{Corbett:2013}, Clock-SI~\cite{Du:2013}, PhysiCS-NMSI~\cite{Tomsic:2016} and CockRoachDB~\cite{cockroachdb}.

\section{Conclusion}
\label{sec:concl}
We have presented \ts{}, a novel geo-replicated key-value store that achieves causal consistency. \ts{} uses hybrid logical/physical clocks and a novel stabilization protocol to achieve better performance, resource utilization and availability than existing approaches.


{\footnotesize \bibliographystyle{acm}
\bibliography{biblio}}
\remove{
\section*{Appendix: Correctness}
\label{sec:corr}
We provide an informal proof that \ts{} implements causal consistency. 
\begin{observation}
\label{prop:ut}
If $X \leadsto Y$, then $X.ut < Y.ut$. 
As a corollary, if $Y.ut \leq X.ut$ then $X \cancel{\leadsto}Y$.
\end{observation}
\begin{proof}
We first show that the observation holds in the case of direct dependency by thread-of-execution or read-from.

Let $c$ be the client that writes $Y$. Upon writing $Y$ at least one of the following two conditions holds: $i)$ $dt_c \geq X.ut$; $ii)$ $USV_c[X.sr] \geq X.ut$.
If $c$ depends on $X$ because $c$ has written $X$, then condition $i)$ holds because of Algorithm~\ref{alg:clnt} Line~\ref{alg:clnt:put:dtc}.  If $c$ depends on $X$ because $c$ has read $X$, there are two cases to consider. If $X$ is local, then condition $i)$ holds because of Algorithm~\ref{alg:clnt} Line~\ref{alg:clnt:get:dtc}. If $X$ is remote, conditions $ii)$ holds because after reading $X$, $USV_c[X.sr] \geq X.ut$ by Algorithm~\ref{alg:srv} Line~\ref{alg:srv:get:get}.
The proposition then holds because Algorithm~\ref{alg:srv} Line~\ref{alg:srv:put:clock} enforces that $Y.ut$ is higher than the update timestamp of any item $c$ depends on.

If $c$ depends on $X$ because of a transitive dependency out of $c$'s thread-of-execution, it means that there is a chain of direct dependencies that leads to a $Z$ such that $X\leadsto \ldots \leadsto Z$ and $c$ has read $Z$. Then, every element in such chain has an update timestamp higher than its predecessor's. Hence, after reading $Z$ at least one between condition $i)$ and $ii)$ holds, leading back to the correctness arguments already exposed for the case of a direct dependency.
\end{proof}

\begin{observation}
\label{prop:ust}
If $X\leadsto Y$, then $Y$ becomes stable only after $X$ does. Namely, for any node $p_n^m$ if $USV_n^m[Y.sr] \geq Y.ut$ then $USV_n^m[X.sr] \geq X.ut$.
\end{observation}
\begin{proof}
Hybrid clocks in \ts{} are monotonically increasing. Similarly, entries in the version vectors of nodes always move forward. The GSV is computed as the aggregate minimum of the version vectors of nodes within a data center. The USV is on its turn computed as the aggregate minimum of the GSV across data centers. 
Therefore, at any node $p_n^m$, $USV_n^m[i]$ can reach the value $t$ only after every item $d$ from the $i$-th data center such that $d.ut \leq t$ has been replicated at all data centers.

If $X.sr == Y.sr$, the proposition holds because $Y.ut > X.ut$ by Observation~\ref{prop:ut}. 
If, instead, $X$ and $Y$ do not share the same originating data center, then there are two cases. Let $c$ be the client that has written $Y$. Then, either $Y$ depends on $X$ because $c$ has read $X$ or because $c$ has read another item $Z$ that depends on $X$.
In the first case, in order for $c$ to read the remote item $X$, then $X$ had been already replicated at all data centers (Algorithm~\ref{alg:srv} Line~\ref{alg:srv:get:get}). Hence, the server $p_n^m$ that has returned $X$ to $c$ at time $T$ was such that $USV_n^m[X.sr] \geq X.ut$. This implies that every node $p_d^k$ in the system at time $T$ was such that $VV_d^k[X.sr] \geq X.ut$. It is, thus, impossible for any $USV$ after time $T$ to include any element created after time $T$ and be such that $USV[X.sr] < X.ut$.

To show that the proposition holds in the remaining case, we can apply a reasoning similar to the one used to demonstrate Observation~\ref{prop:ut} in case of transitive dependencies.
\end{proof}

\begin{observation}Let $X, Y$ be two local items such that $X \leadsto Y$. Then $X.DV \leq Y.DV$.
\label{prop:dv}
\end{observation}
\begin{proof}
Let $c$ be the client that writes $Y$ in data center $m$. If $c$ has also written $X$, then the proposition holds because $dt_c$ and the entries of $USV_c$, used to build the dependency vectors of $X$ and $Y$, are monotonically increasing (Algorithm~\ref{alg:clnt} Lines~\ref{alg:clnt:get:usv}, \ref{alg:clnt:get:dtc}, \ref{alg:clnt:put:dtc} and ALgorithm~\ref{alg:srv} Line~\ref{alg:srv:put:clock}).
If $c$ reads $X$ from node $p_n^m$, then $USV_c[i] \geq X.DV[i], \forall i=0,\ldots, M, i\neq m$. In fact, $X$ can only depend on remote items that are stable in the whole system. Also, it is $dt_c \geq X.ut$ because of Algorithm~\ref{alg:clnt} Line~\ref{alg:clnt:get:dtc}. Since $X.DV[m] < X.ut$ by Observation~\ref{prop:ut} and $Y.DV[m] = dt_c$, the proposition follows.

We now show that the proposition holds in case there is a chain of direct dependencies $X\leadsto Z \leadsto Y$ and $Y$ reads $Z$. If $X\leadsto Z$ there are two cases. If $Z$ is local, then we apply the previous reasoning to show that $Y.DV \geq Z.DV \geq X.DV$. If $Z$ is remote, then Observation~\ref{prop:ust} enforces that, upon reading $Z$ locally, the entries of the $USV$ returned to the client are higher than the timestamp of any dependency of $Z$, i.e, $USV_c \geq X.DV$. Hence, $Y.DV \geq USV_c \geq X.DV$.

Any dependency chain is a combination of the three aforementioned cases. So, to show that the proposition holds in the general case, we can apply the same iterative reasoning described to demonstrate previous prepositions.
\end{proof}

\begin{proposition}
\label{prop:get}
Let $X\leadsto Y$. If client $c$ reads $Y$, any subsequent \texttt{get(x)} operation issued by $c$ will return $X': X'\cancel{\leadsto}X$.
\end{proposition}
\begin{proof}
We show that if $c$ reads $Y$, then $X$ is visible to $c$. Hence, since a server always return the available visible version of a key with the highest timestamp (Algorithm~\ref{alg:srv} Line~\ref{alg:srv:get:get}), if $X$ is visible to $c$ and $c$ wants to read $x$, either $X$ is returned or a $X'': X''.ut > X.ut$. Such $X''$ cannot depend on $X$ by Observation~\ref{prop:ut}.

We first show that the proposition holds in case $Y$ depends on $X$ directly, i.e., not transitively. We note $c'$ the client that has written $Y$ after writing (thread-of-execution) or reading (read-from) $X$. 

If $c$ and $c'$ are in the same data center, then in both cases the proposition holds because updates are immediately visible in the originating data center and because by Observation~\ref{prop:ut}, if $X' \leadsto X$, then $X'.ut < X.ut$. 

If $c'$ is in a different data center $DC_i$ from $c$, there are two cases to consider: $i)$ $c'$ has written $X$ and then $Y$; or $c'$ has read $X$ and then written $Y$. In both cases, by Observation~\ref{prop:ust}, $Y$ becomes stable only after $X$ is. Therefore, after $c$ reads $Y$, $USV_c[X.sr] \geq X.ut$ (Algorithm~\ref{alg:clnt} Line~\ref{alg:clnt:get:usv}). Let $p_n^m$ the server responsible for serving the later $get(x)$ operation issued by $c$. Then, $USV_n^m[X.sr] \geq X.ut$ by Algorithm~\ref{alg:srv} Line~\ref{alg:srv:get:usv}, making $X$ visible to $c$.

We now show that the proposition holds in case $Y$ transitively depends on $X$. In case of transitive dependency, there exists an item $Z$ (corresponding to an arbitrary key) such that $X\leadsto Z \leadsto Y$. In case $X\leadsto Z$ directly and $Z \leadsto Y$ directly, then the proposition holds because if $c$ reads $Y$, then any subsequent get operation would be served by a snapshot that includes $Z$. Since $X\leadsto Z$, this also means that any snapshot including $Z$ includes $X$ too, thus proving the proposition.

If $X\leadsto Y$ because of a longer chain of dependencies, we note that such chain is composed of only direct dependencies. Hence, we can iteratively apply the reasoning just described to show that the proposition holds.
\end{proof}

\begin{proposition}
\label{prop:tx}
\ts{} implements causally read-only transactions as defined in Section~\ref{sec:def:model}.
\end{proposition}
\begin{proof}
We first show that if $c$ reads $Y$ in a transaction, then $Y$ is causally consistent with $c$'s history. 
Upon issuing a read-only transaction, $USV_c$ and $dt_c$ are communicated to the transaction coordinator $p_n^m$. Any remote item $Z$ read by $c$ is such that $Z.ut \leq USV_c[Z.sr]$. Since the transaction vector $TV$ is such that $TV \geq USV_c$, $Z$ is visible within the scope of the transaction according to what already shown for the case of $get$ operations. 

The visibility of local items is, instead, managed differently from the case of the simple $get$. Any local item $L$ written or read by $c$ is such that $L.DV \leq TV$. In fact, $TV[i] = USV_n^m[i] \geq USV_c[i] \geq L.DV[i], i=0\ldots M, i \neq j$ and $TV[j] \geq dt_c$. Hence,  the transaction snapshot includes also all local items on which $c$ directly or transitively depends on.

We now show that if a read-only transaction returns $X, Y : X \leadsto Y$, then $\nexists X': X\leadsto X'\leadsto Y$. Specifically, we show that if such $X'$ existed, then it would be in the snapshot visible to the transaction. Therefore, since $X'.ut > X.ut$ by Observation~\ref{prop:ut}, returning $Y, X$ instead of $Y, X'$ would not be possible. We consider four different cases, corresponding to the possible local/remote combinations of the source replicas of $X'$ and $Y$.
$i)\ Y\ local,\ X'\ local.$ By Observation~\ref{prop:dv}, $Y.DV \geq X'.DV$. By Algorithm~\ref{alg:srv} Line~\ref{alg:srv:slice:loc}, $TV \geq Y.DV$. Then, it follows that $X'.DV \leq TV$ and hence $X'$ is in the snapshot.
$ii)\ Y\ local,\ X'\ remote.$ By Observation~\ref{prop:dv}, $Y.DV[X.sr] \geq X.ut$, and by Algorithm~\ref{alg:srv} Line~\ref{alg:srv:slice:loc} $TV \geq Y.DV$. Hence $TV[X'.sr] \geq X'.ut$, i.e., $X'$ is in the snapshot.
$iii)\ Y\ remote,\ X'\ remote.$ For $Y$ to be in the snapshot, it is $TV[Y.sr] \geq Y.ut$. Then, since $X'\leadsto Y$, by Observation~\ref{prop:ust} $X'$ is visible in any snapshot that includes $Y$.
$iv)\ Y\ remote,\ X'\ local.$ For $Y$ to depend on $X$ it means that $X'$ and all its remote dependencies had already been fully replicated when $Y$ has been created. Hence, by Observation~\ref{prop:ust}, $X'.DV \leq TV$ and hence it is visible.

{\color{blue}{We finally note that the proposed implementation of read-only transactions allows \ts{} to return a local (from data center $m$) $X'$ whose update timestamp falls outside the boundaries of the transaction snapshot as long as all the dependencies of $X'$ are within said snapshot. If one wanted to disallow this behavior, it would be sufficient to enforce in Algorithm~\ref{alg:srv} Line~\ref{alg:srv:slice:loc} that any local item $Z$ returned by a transaction also complies with the condition $Z.ut < TV[m]$.}}
\end{proof}
}

\end{document}